%% file: elsarticle-template-num-names.tex
\documentclass[final,5p,times,twocolumn]{elsarticle}
\usepackage{amssymb}
\usepackage{amsmath}
\usepackage{wasysym}
\usepackage{graphicx} % Required for inserting images
\usepackage{adjustbox}
\usepackage[hidelinks]{hyperref}
\usepackage{fix-cm}
\usepackage{multirow}
\usepackage{amsfonts}
\usepackage[table]{xcolor}
\usepackage{xcolor}

\usepackage{booktabs}
\usepackage{pifont} 
\usepackage{color,soul}
\usepackage{enumitem}
\usepackage{tikz}
\usepackage{lineno}
\newcommand{\supplementary}{
  \setcounter{section}{0}
  \renewcommand{\thesection}{\Alph{section}}
  \section*{Supplementary Material}
  \addcontentsline{toc}{section}{Supplementary Material}
}
\journal{Nuclear Physics B}
\usepackage{xcolor}
\newcommand{\revised}[1]{\textcolor{black}{#1}}
\begin{document}

\begin{frontmatter}



\title{A Taxonomy of Attacks and Defenses in Split Learning} %% Article title
\author[1]{Aqsa Shabbir\fnref{equal}}
\author[2]{Halil İbrahim Kanpak\fnref{equal}}
\author[2]{Alptekin Küpçü\corref{cor1}}
\author[1]{Sinem Sav\corref{cor1}}

\address[1]{Bilkent University, Ankara, Türkiye}
\address[2]{Koç University, İstanbul, Türkiye}

\fntext[equal]{Equal contribution.}
\cortext[cor1]{Corresponding author.\\
\emph{E-mail addresses:} 
\href{mailto:akupcu@ku.edu.tr}{akupcu@ku.edu.tr} (A. Küpçü), 
\href{mailto:sinem.sav@cs.bilkent.edu.tr}{sinem.sav@cs.bilkent.edu.tr} (S. Sav).}


% \author{} %% Author name

% \affiliation{organization={},%Department and Organization

%% Abstract
\begin{abstract}
Split Learning (SL) has emerged as a promising paradigm for distributed deep learning, allowing resource-constrained clients to offload portions of their model computation to servers while maintaining collaborative learning. However, recent research has demonstrated that SL remains vulnerable to a range of privacy and security threats, including information leakage, model inversion, and adversarial attacks. While various defense mechanisms have been proposed, a systematic understanding of the attack landscape and corresponding countermeasures is still lacking. In this study, we present a comprehensive taxonomy of attacks and defenses in SL, categorizing them along three key dimensions: employed strategies, constraints, and effectiveness. Furthermore, we identify key open challenges and research gaps in SL based on our systematization, highlighting potential future directions.
\end{abstract}


% \begin{highlights}
% \end{highlights}

%% Keywords

\end{frontmatter}


\input{introduction}

\input{background}
\input{threatsattacks}
\input{defensemechanisms}

\input{takeaways}
\input{related}

\input{conclusion}
\bibliographystyle{abbrv}
\bibliography{reference}
\newpage
\supplementary
\input{supplementary}
\end{document}

%% file: introduction.tex
\section{Introduction}

The increasing adoption of Machine Learning as a Service (MLaaS) has revolutionized AI-driven applications across domains such as healthcare, finance, and IoT. However, ML outsourcing raises privacy concerns, as sensitive user data is often processed on external or untrusted servers~\cite{ryan2011cloud}. To address this challenge, privacy-preserving machine learning (PPML) techniques have been developed, enabling collaborative learning while ensuring data confidentiality.

Among various collaborative methods, Split Learning (SL) emerged as a promising framework that partitions deep learning models between clients and servers~\cite{gupta2018split}. SL enables clients to process several layers of a neural network locally while delegating the remaining layers to a server. This design reduces the computational burden on resource-constrained client devices while keeping raw data in the local premises of the client~\cite {vepakomma2018split}. Over time, SL has evolved into different architectures, including Vanilla Split Learning (VanSL)~\cite{gupta2018split}, U-shaped Split Learning (USL)~\cite{gupta2018split}, No-Label Split Learning (NLSL)~\cite{li2021label}, Hybrid Split Learning (Hybrid-SL)~\cite{gao2022combined}, and Multi-hop Split Learning (MHSL)~\cite{vepakomma2018split}, each optimizing trade-offs between performance, scalability, and privacy.

Despite its architectural advantages for resource-constrained clients and inherent data locality~\cite{gupta2018split,vepakomma2018split}, SL is vulnerable to several security and privacy risks. Data reconstruction attacks exploit smashed data and gradients to infer private inputs, while label inference attacks analyze gradient updates to recover class labels~\cite{Unleashing_tiger,fu2022label}. Property inference attacks extract sensitive attributes without full data reconstruction~\cite{mao2023securesplit}, and model inversion~\cite{erdougan2022unsplit} aims to recover input samples from model activations. Additionally, backdoor and poisoning attacks~\cite{bai2023villain,yu2024chronic} manipulate the training pipeline to implant adversarial behavior. These vulnerabilities expose critical gaps in the privacy guarantees of SL, necessitating robust defense mechanisms to mitigate potential threats. \revised{However, before effective defenses can be designed and evaluated, a systematic understanding of the \textit{attack landscape} itself is required. We build this foundation by addressing these core research questions regarding \textit{attacks} against SL:}

\revised{\textit{RQ1:} \textit{How can adversarial objectives targeting SL be systematically taxonomized? Which architectural vulnerabilities in SL implementations serve as primary attack vectors?}}

\revised{\textit{RQ2:} \textit{What are the fundamental characteristics and underlying principles of different SL attack mechanisms (e.g., feature-space hijacking, model inversion, gradient-based inference, embedding poisoning) that dictate their generalizability across SL variants and data domains, their potential for stealth, and their ability to bypass defenses?}}

\revised{\textit{RQ3:} \textit{Under what operational constraints and adversarial assumptions (regarding knowledge, access, capabilities, and collusion scenarios) do attacks against SL demonstrate measurable effectiveness? How can their quantifiable impact on privacy and model integrity be systematically evaluated?}}

To counter these threats, researchers have explored various privacy-preserving techniques as defense strategies, including DP to perturb gradients~\cite{yang2022differentially}, homomorphic encryption (HE) for encrypted computation to prevent unauthorized data access~\cite{splitHE}, and communication compression to reduce information leakage \cite{pham2022binarizing}. While each method strengthens SL security, it also introduces trade-offs in computational overhead, scalability, and model performance. \revised{Therefore, a clear understanding of the defense landscape, its capabilities, limitations, and evaluation is equally critical. Consequently, this paper also addresses these research questions concerning \textit{defenses} in SL:}

\revised{\textit{RQ4:} \textit{What is the taxonomy of defense techniques against predefined attack scenarios/ adversarial objectives tailored for SL? Wat are the common techniques and tools being employed?}}

\revised{\textit{RQ5:} \textit{Which defenses are suitable for a specific task? What are the constraints and conditions implied by these defenses? Can these techniques be optimized? Are these techniques suitable for real-world applications?}} 

\revised{\textit{RQ6:} \textit{How do defense techniques aim for success among different scenarios in terms of confidentiality or integrity of training? What constitutes a successful defense mechanism?}}

\revised{This paper provides a comprehensive SL analysis by systematically categorizing attack vectors and defense strategies based on these research questions. We propose taxonomies for attacks and defenses, evaluating their methodology, constraints, and effectiveness. We highlight key trade-offs, our observations, and open challenges derived from this systematization, paving the way for future advancements in SL privacy and robustness. %By consolidating recent developments, this work offers a structured perspective on current threats and countermeasures, enabling researchers and practitioners to build more resilient and privacy-preserving SL-based systems.
}

%The rest of the paper is structured as follows: Section~\ref{sec:method} outlines our review scope and systematization methodology. In Section~\ref{sec:background}, we provide background on split learning. Section~\ref{sec: Threats and Attack Scenarios} \revised{addresses RQ1-RQ3 by presenting our taxonomy and analysis of attacks against SL, followed by Section~\ref{sec:Defense Mechanisms}, which addresses RQ4-RQ6 by classifying and analyzing defense mechanisms. Section~\ref{sec:takeaways} summarizes our key findings and identifies gaps across all RQs. Section~\ref{sec:Open Research Directions} outlines promising directions for future research.} Finally, Section~\ref{sec:related} provides the related work.

\section{Review Scope and Methodology}\label{sec:method}

% This paper presents a Systematization of Knowledge (SoK) on Split Learning (SL): a distributed machine learning paradigm that enhances privacy while distributing computational workloads between clients and a server(s). %While SL offers advantages over Federated Learning (FL) by reducing communication overhead and enabling efficient multi-party training, it also introduces significant security and privacy risks. 
We systematically categorize SL architectures, identify key attack surfaces —including data reconstruction, feature leakage, and adversarial interference— and evaluate existing defense mechanisms such as differential privacy, homomorphic encryption, and adversarial training. By consolidating insights from prior research, we provide a structured framework for assessing privacy risks, defense effectiveness, and security trade-offs, laying the foundation for future advancements in PPML. %Below, we detail our search and systematization methodology. 

\par\noindent\textbf{Search Methodology.} To ensure a comprehensive and representative collection of research on SL, we systematically gathered papers from major academic databases, including Google Scholar, IEEE Xplore, ACM Digital Library, arXiv, and SpringerLink. The search was conducted in two stages: an automated search using predefined keyword queries, followed by a snowballing approach to refine and expand the selection of relevant studies. We used keyword combinations to capture diverse SL research, including privacy, attack surfaces, and defenses. Example keywords included a combination of “\textit{split learning}”, “\textit{privacy-preserving}”, “\textit{split neural networks}”, “\textit{privacy attacks on split learning}”, “\textit{homomorphic encryption based split Learning}”, “\textit{differential privacy based split learning}”, and “\textit{inference attacks in split learning}”. These queries targeted research addressing both theoretical and practical aspects of SL, ensuring coverage of multiple perspectives on its security vulnerabilities and mitigation strategies. Then, we employed a snowballing approach, where we examined the references of key papers to identify prior work. Expanding this approach in both directions, we also examined papers that have cited the ones we identified, allowing us to trace the evolution of knowledge in the field. %Additionally, we selected papers that were either published in a conference/journal or cited by such publications. 
% We review \hl{22 papers} for the attacks and \hl{23 papers} for the defense mechanisms in total.
%During our search, we combined multiple dimensions of SL (e.g., ``single-client single-server'' OR ``homomorphic encryption based split learning'' OR ``malicious split learning'') to systematically locate relevant studies.  

After identifying works across these different feature aspects, we aligned and classified them to develop our overall taxonomy for SL systematization. During the filtering process, each identified paper was first assessed based on its abstract, followed by a thorough review of its methodology and experimental evaluation if the abstract indicated the use of a relevant technique. Papers that did not explicitly reference SL, such as those focused on general distributed machine learning, were excluded. We then classified each paper as comprehensively as possible, considering multiple aspects such as the defense mechanisms it proposes in case of an attack paper and the attacks it addresses in case of a defense paper, the assumed threat model, and the server-client setup. 
\par\noindent\textbf{Systematization Methodology.} For our systematization, we begin by reviewing SL architectures to establish a foundational understanding of the landscape. We then categorized the collected papers into two primary groups: attack papers and defense papers. Within each group, we employ a structured classification approach based on three key dimensions:
\begin{enumerate}
    \item Strategies: The method or strategies used to conduct attacks or implement defenses.
    \item Constraints: The assumptions, limitations, and requirements under which the attacks or defenses operate.
    \item Effectiveness: The impact and success rate of attacks, as well as the robustness and trade-offs of defense mechanisms.
\end{enumerate}

\begin{figure}[htbp] 
\centering
\includegraphics[width=0.48\textwidth]{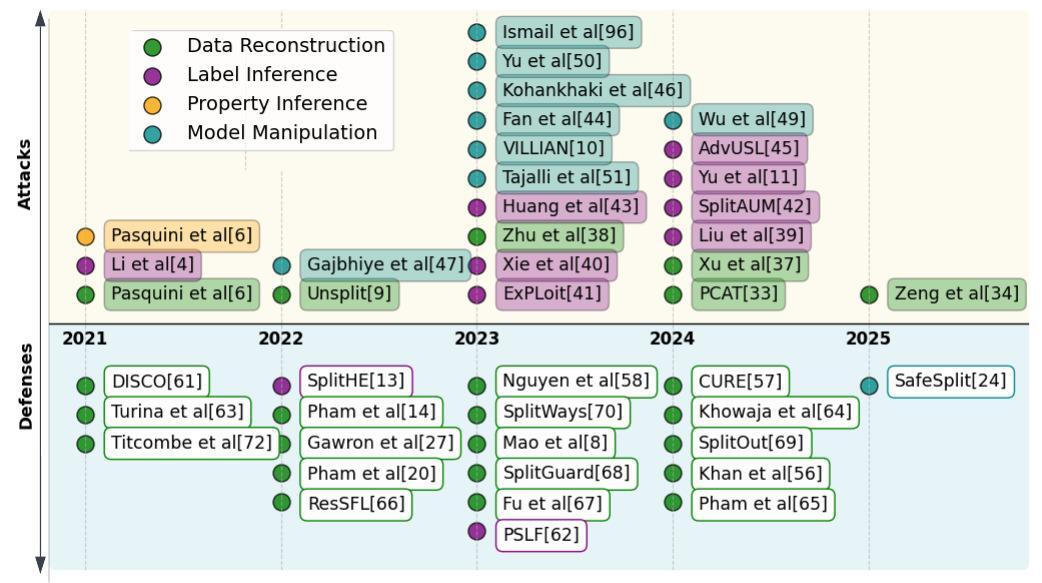}
\caption{SL attack and defense timeline (2021–2025), with attacks on the top half and defenses on the bottom half. Attack and defense types are categorized into four groups, as shown in the legend. Attacks are represented with a colored background, while the corresponding defenses share the same color but are displayed without a background. }
% \Description{SL attack and defense timeline (2021–2025)}
\label{fig:Attack and Defense studies}
    \vspace{-1em}
\end{figure}

Finally, we present a timeline in Figure~\ref{fig:Attack and Defense studies} that compiles our surveyed literature on both attacks and defenses. It illustrates the chronological progression and dominant strategies.

%% file: background.tex
\section{\revised{Split Learning (SL)}}\label{sec:background}
We provide an overview of split learning (SL), its variants, and key concepts. We provide the frequently used notations and abbreviations in Table~\ref{tab:notations}. %, Supplementary Material~\ref{ref:notations}. 
% We discuss how SL addresses privacy concerns, reduces computational overhead, and supports collaborative learning across distributed datasets.  

\begin{table}[ht]
    \centering
    \caption{Key Notations and Abbreviations}
    \label{tab:notations}
    \begin{tabular}{ll}
    \toprule
    \textbf{Notation/Abbreviation} & \textbf{Description} \\
    \midrule
    % Core SL Components
    $f_c$             & Client-side network model part \\
    $f_s$             & Server-side network model part \\
    $x$               & Raw input data (client-side) \\
    $y$               & True label (often client-side) \\
    $z_c$             & Smashed data \\
    $\nabla_{z_c}$    & Gradient sent from server to client \\
    $\hat{y}$         & Model's final output/prediction \\

    % Attack-Related
    $\tilde{x}$       & Reconstructed input data \\
    $\tilde{y}$       & Inferred label \\

    % SL Variants
    VanSL             & Vanilla Split Learning \\
    USL               & U-Shaped Split Learning \\
    SFL               & Split Federated Learning \\
    HSL               & Horizontal Split Learning \\
    VSL               & Vertical Split Learning \\ % 

    % Architectures
    SCSL              & Single-Client Split Learning \\
    MCSL              & Multi-Client Split Learning \\
    SSSL               & Single-Server Split Learning \\
    MSSL               & Multi-Server Split Learning \\

    % Threat Models
    SHS               & Semi-Honest Server \\
    MS                & Malicious Server \\
    MC                & Malicious Client \\
    SHC               & Semi-Honest Client \\

    % Common Defense/Related Tech
    FL                & Federated Learning \\
    HE                & Homomorphic Encryption \\
    DP                & Differential Privacy \\

    \bottomrule
    \end{tabular}
\end{table}

\subsection{Split Learning and Its Variants}
\label{sec: Split Learning Variants}
\begin{figure*}
  \centering  \includegraphics[width=0.75\textwidth]{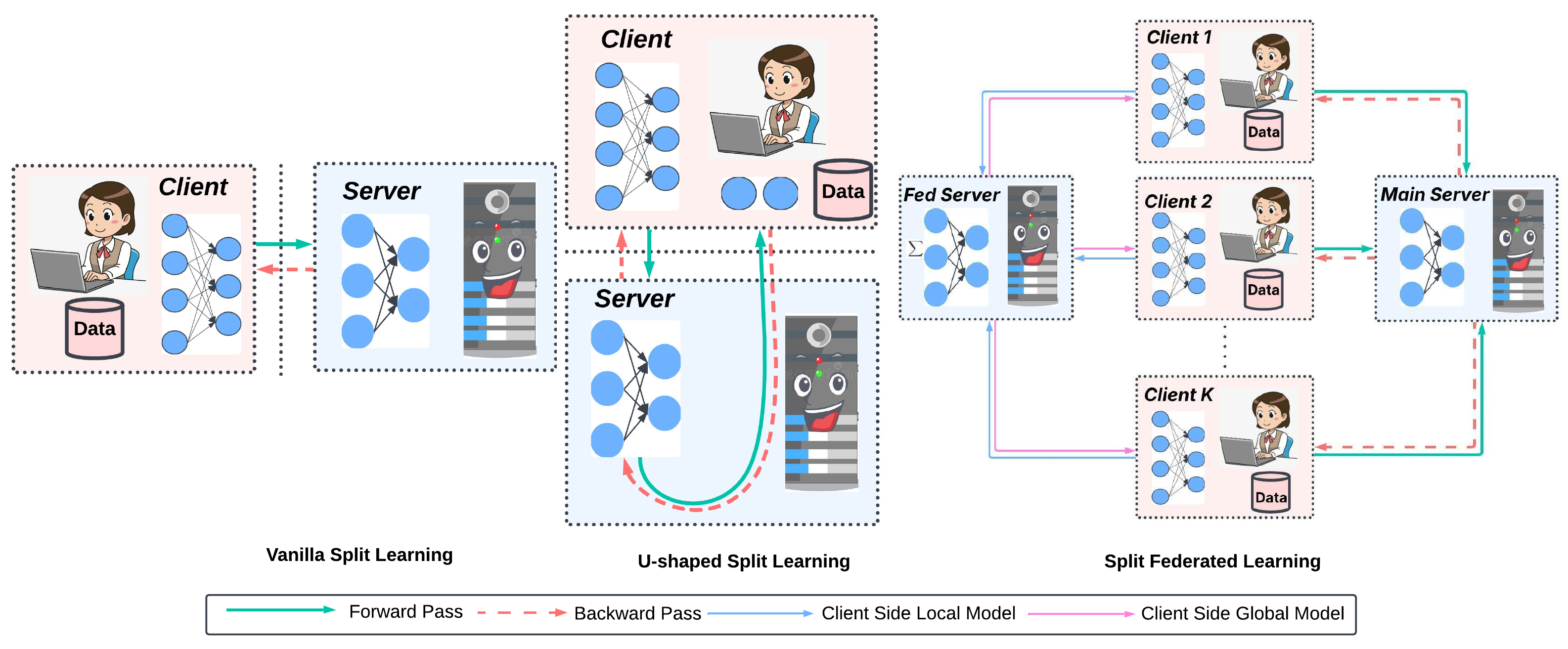} \caption{Overview of three key SL variants: Vanilla Split Learning, U-shaped Split Learning, and Split Federated Learning. The figure highlights their architectural distinctions, client-server roles, and the flow of forward and backward passes in training.}
  % \Description{A diagram showing split learning styles.}
  \label{fig:Split Learning}
\end{figure*}

SL~\cite{gupta2018split} is a distributed machine learning paradigm where the computation of a deep learning model is split between clients and servers. It has evolved into various forms to address the diverse needs of collaborative machine learning. The key variants include Vanilla SL~\cite{gupta2018split}, U-shaped SL~\cite{gupta2018split}, no-label SL~\cite{li2021label}, Hybrid SL~\cite{thapa2022splitfed}, and Multi-hop SL~\cite{vepakomma2018split}.  %These variants differ based on the distribution of model computation and their suitability for specific computational and privacy-preserving challenges. 
Additionally, the effectiveness of SL can be influenced by the choice of data partitioning schemes, namely horizontal partitioning and vertical partitioning, which dictate how data is distributed among participants. We detail these variants below and illustrate the most commonly used ones in Figure~\ref{fig:Split Learning}.
\par\noindent\textbf{Vanilla Split Learning (VanSL):}
\label{VanSL}
VanSL is the simplest form of SL, where the client processes the \textit{initial layers} $(f_c)$, and the server processes the \textit{remaining layers} $(f_s)$. The client uses its private data $x$ to compute the \textit{intermediate activations} $z_c = f_c(x)$, which are transmitted to the server. The server completes the \textit{forward pass} by calculating the output $\hat{y} = f_s(z_c)$. During backpropagation $(f')$, the server computes the gradient of the loss function $L$ with respect to $z_c$, denoted as $\nabla_{z_c} = \frac{\partial L}{\partial z_c}$, and sends it back to the client. The client then updates the parameters of $f_c$ using its local gradients. This collaborative process ensures the client retains $x$. % while enabling end-to-end model training.
\par\noindent\textbf{U-shaped Split Learning (USL):}\label{USL}
USL divides the model into three segments: (i) \textit{initial layers} $f_c$ processed by the client, (ii) \textit{middle layers} $f_s$ handled by the server, and (iii) \textit{remaining layers} $f_{c\_r}$ completed by the client. The client begins by processing $x$ through $f_c$ to compute $z_c$ and sends $z_c$ to the server for further computation. The server processes $z_c$ through $f_s$ to compute $z_s$ and returns $z_s$ to the client. The client completes the forward pass by applying $f_{c\_r}$ to $z_s$, resulting in the final output $\hat{y}$. The backpropagation process follows the same sequence with a similar approach, in reverse.
% During backpropagation $f'$, the gradient $\nabla_{\hat{y}}$ is first computed on the client and propagated through $f_{c\_r}$, updating parameters locally. The client then computes $\nabla_{z_s}$ and sends it to the server. The server uses $\nabla_{z_s}$ to backpropagate through $f_s$,  updating its parameters. Finally, the server computes $\nabla_{z_c}$ and sends it back to the client, allowing the client to update the parameters of $f_c$. 
This way, USL ensures that $x$ and $\hat{y}$ remain at the client while enabling end-to-end training. % of the split model.

\par\noindent\textbf{Hybrid Split Learning:}\label{HybSL}
Hybrid SL integrates SL with other collaborative learning paradigms, such as federated learning (FL) to enhance privacy, scalability, and efficiency. For instance, combining SL with FL, Split Federated Learning (SFL) allows clients to train parts of the model collaboratively while leveraging federated learning aggregation techniques. Similarly, incorporating HE ensures that smashed data exchanged between clients and servers remains encrypted, further strengthening privacy. It enables solutions for scenarios involving heterogeneous clients and large-scale data distributions. However, integrating additional mechanisms often introduces computational and communication overhead, making efficient design and implementation critical for practical use.

We present less commonly adopted SL variants, i.e., \textbf{multi-hop} and \textbf{no-label SL}, in Supplementary Material~\ref{slvariants}. Finally, unlike Federated Learning (FL)~\cite{zhao2018federated}, where the entire model is trained locally on each client and the gradients are aggregated by a server, SL partitions the model, delegating the more computationally intensive layers to the server. This makes SL~\cite{vepakomma2018split} ideal for resource-constrained clients, who wish to outsource most of the computationally-intensive training procedure.

\subsection{Data Partitioning Schemes} \label{data partitioning}
Data partitioning in SL determines how data is distributed among clients, impacting both the training process and privacy assurances. In \textbf{Horizontal Split Learning (HSL)}~\cite{li2024split,vepakomma2018split}, clients handle data with identical feature spaces but different samples; for instance, multiple hospitals with similar patient data but distinct patient groups can collaboratively train a model by sharing intermediate activations without exposing raw data. Conversely, \textbf{Vertical Split Learning (VSL)}~\cite{vepakomma2018split,joshi2022performance} is applicable when clients possess complementary features for the samples; each client processes its unique features, and a server combines these embeddings to facilitate joint learning, such as a bank and an e-commerce platform sharing distinct user attributes to jointly develop a credit scoring model.

% Data partitioning schemes in SL define how data is distributed among clients, influencing the training process and privacy guarantees. Below, we detail the main partitioning strategies: 
% \par\noindent\textbf{Horizontal Split Learning:} Horizontal SL (HSL)~\cite{li2024split,vepakomma2018split} applies to scenarios where multiple clients possess data with the same feature space but different data samples. In this approach, each client processes its portion of the data independently, contributing to a collaborative learning process. For example, in a healthcare setting, multiple hospitals with similar patient features but distinct patient populations can collaboratively train a model by sharing intermediate activations without exposing raw data.
% \par\noindent\textbf{Vertical Split Learning:} Vertical SL (VSL)~\cite{vepakomma2018split,joshi2022performance} is suited for cases where clients hold complementary feature spaces for the same set of samples. Each client processes its unique features, and the server combines these embeddings to enable collaborative learning. For example, in a financial setting, a bank and an e-commerce platform share user data with distinct attributes (e.g., transaction history and purchase behavior) to jointly train a model for credit scoring while keeping their respective data private.

% \end{enumerate}
% \begin{figure*}
%   \centering  \includegraphics[width=\textwidth]{Data Partitioning.jpg} \caption{Data Partitioning Schemes}  
%   \label{fig:Data Partitioning}
% \end{figure*}
\subsection{Client-Server Participation Models}
\label{sec: Split Learning Architectures}
%Here, we define SL’s client-server participation models, detailing the collaboration dynamics between multiple clients and servers.
% It determines how the model is split, how data flows between the client and the server, and how updates are shared.
\par\noindent\textbf{Single Client Split Learning (SCSL)}
In SCSL~\cite{vepakomma2019reducing}, training involves one client and a server, making it ideal for a single organization, such as a hospital or research institute, that seeks to preserve privacy while offloading the heavy computation to a server. Although it simplifies communication and coordination, clients still need sufficient resources to process their local model portions.

\par\noindent\textbf{Multi Client Split Learning (MCSL)}
MCSL~\cite{pham2022split} keeps the forward and backward propagation split between multiple clients and a server. Clients only send smashed data to the server, and the server processes the deeper layers of the model and updates all clients without requiring them to maintain a full model locally.

\par\noindent\textbf{Single Server Split Learning (SSSL)}
In SSSL~\cite{vepakomma2018split}, all clients interact with a single server. Clients send intermediate activations to the server, which processes the remaining model layers and computes gradients. The server then returns these gradients to clients. Unlike MCSL, where multiple clients interact with multiple servers, SSSL uses a single-server architecture with all clients relying on one server for training. This setup is advantageous for organizations with centralized infrastructure capable of handling high computational demands. It simplifies system architecture and reduces coordination overhead but may introduce scalability limitations as the number of clients increases.

\par\noindent\textbf{Multi-Server Split Learning (MSSL)}
MSSL~\cite{pham2022split} divides the model computation across multiple servers, allowing better scalability and resource utilization. Each client processes the initial model layers locally and sends the smashed data to one or more servers. The servers handle subsequent computations, often partitioning the model layers or tasks among themselves, and collaboratively complete forward and backward passes. MSSL is ideal for large-scale deployments where a single server cannot handle the computation or communication demands of multiple clients. It enhances system scalability and fault tolerance but requires efficient communication and synchronization between servers.

% \subsection{The Scope of the Study}
% This paper presents a Systematization of Knowledge (SoK) on SL, a distributed machine learning paradigm designed to enhance data privacy while distributing computational workloads between clients and a server. While SL offers a promising alternative to FL by reducing communication overhead and enabling efficient model training across multiple parties, it also introduces new security and privacy risks. This study systematically categorizes existing SL architectures and identifies the vulnerabilities associated with each approach. It further explores various attack surfaces in SL, such as data reconstruction, feature leakage, and adversarial interference. In addition, we review existing defense mechanisms, including differential privacy, homomorphic encryption, and adversarial training, and evaluate their effectiveness in mitigating these threats. By consolidating insights from prior research and highlighting key challenges, this study aims to provide a comprehensive framework for assessing the security, privacy, and efficiency trade-offs in SL, paving the way for future research in privacy-preserving machine learning.
\revised{
\subsection{Positioning SL within the PPML Landscape}
%Despite its widely discussed vulnerabilities \ref{sec: Threats and Attack Scenarios}, Split Learning retains practical value in certain deployment scenarios, especially when compared with other privacy-preserving machine learning (PPML) paradigms. 
While SL exposes intermediate representations to the server, leading to attacks such as model inversion, label inference, and feature space hijacking, other PPML methods, such as Federated Learning (FL) and Secure Aggregation, are not immune to similar threats. For instance, in FL, gradients shared by clients can be exploited to reconstruct raw inputs, even without access to the model architecture or full training data~\cite{zhu2019dlg}. Similarly, Secure Aggregation protocols may be circumvented when adversaries collude or exploit partial gradient information to infer sensitive data~\cite{Hitaj2017,Nasr2019Comprehensive}. These observations suggest that the underlying vulnerability (exposure of sensitive gradients or activations) is a broader issue in collaborative learning, not one unique to SL.
Nonetheless, its advantages and disadvantages must be weighed carefully.}

\revised{
One of SL’s key strengths is offloading the client's computational burden to the server. In standard setups such as Vanilla SL or U-shaped SL, the client only processes a few initial layers of the model and never needs to store or train the full network~\cite{gupta2018split,vepakomma2018split}. This makes SL particularly suitable for resource-constrained devices such as smartphones, wearables, or embedded IoT systems in domains like digital health, where data confidentiality is crucial and local compute capacity is limited. %Another advantage is that raw input data remains on the client device, thereby avoiding direct exposure, even though adversaries may still infer properties or reconstruct inputs from intermediate representations \cite{Unleashing_tiger, erdogan2024splitout}.
}

\revised{
However, SL also has its limitations: It lacks robust mechanisms for verifying the integrity of the training process, leaving it vulnerable to model manipulation, data poisoning, and backdoor attacks, especially in multi-client settings where colluding parties can exploit sequential training dynamics~\cite{bai2023villain,li2021label,safeSplit2025rieger}. Moreover, SL’s training is often serialized, particularly in multi-client architectures, leading to communication bottlenecks and slower convergence compared to methods that allow concurrent client updates~\cite{pham2022split}.}

\revised{Given these trade-offs, SL is most appropriate in scenarios where deep models are essential, client computational power is constrained, and data sensitivity is high. Examples include natural language processing tasks involving transformer-based models or clinical settings with privacy-sensitive patient records~\cite{roth2022splitunet}. 
%Recent implementations of SL in domains such as brain tumor segmentation \cite{roth2022splitunet} or question-answering with large language models demonstrate this contextual suitability. 
Poirot et al.~\cite{poirot2019split} demonstrated the effectiveness of SL in healthcare across disparate datasets while preserving privacy. 
Wang et al.~\cite{wang2024stitch}  explored Stitch-able SL for multi-UAV systems, showcasing its ability to handle device instability and model heterogeneity in complex tasks. 
Li et al.~\cite{li2022federated} applied Split Learning to BERT-based models, achieving communication efficiency and privacy preservation in decentralized text classification.}

%% file: threatsattacks.tex
\section{Attacks on Split Learning}
\label{sec: Threats and Attack Scenarios}
SL vulnerabilities arise from shared intermediate activations, gradients, and model states, exposing potential attack surfaces. This section categorizes these threats across three dimensions: \emph{attack strategies} (Section~\ref{sec: Attack Tecniques Employed}), \emph{operational constraints} (Section~\ref{sec: Attack Constraints}), and \emph{attack effectiveness} (Section~\ref{sec: Attack Effectiveness}). Finally, we provide a summary of the surveyed literature in Table~\ref{tab:general table}, Supplementary Material~\ref{sec:supplementaryTable}.
% \begin{figure}
%   \centering  \includegraphics[width=\columnwidth]{Attack/SL Attack Papers.png} \caption{Attack studies}  
%   \label{fig:Attack studies}
% \end{figure}
\begin{figure}[htbp]
  \centering  \includegraphics[width=0.47\textwidth]{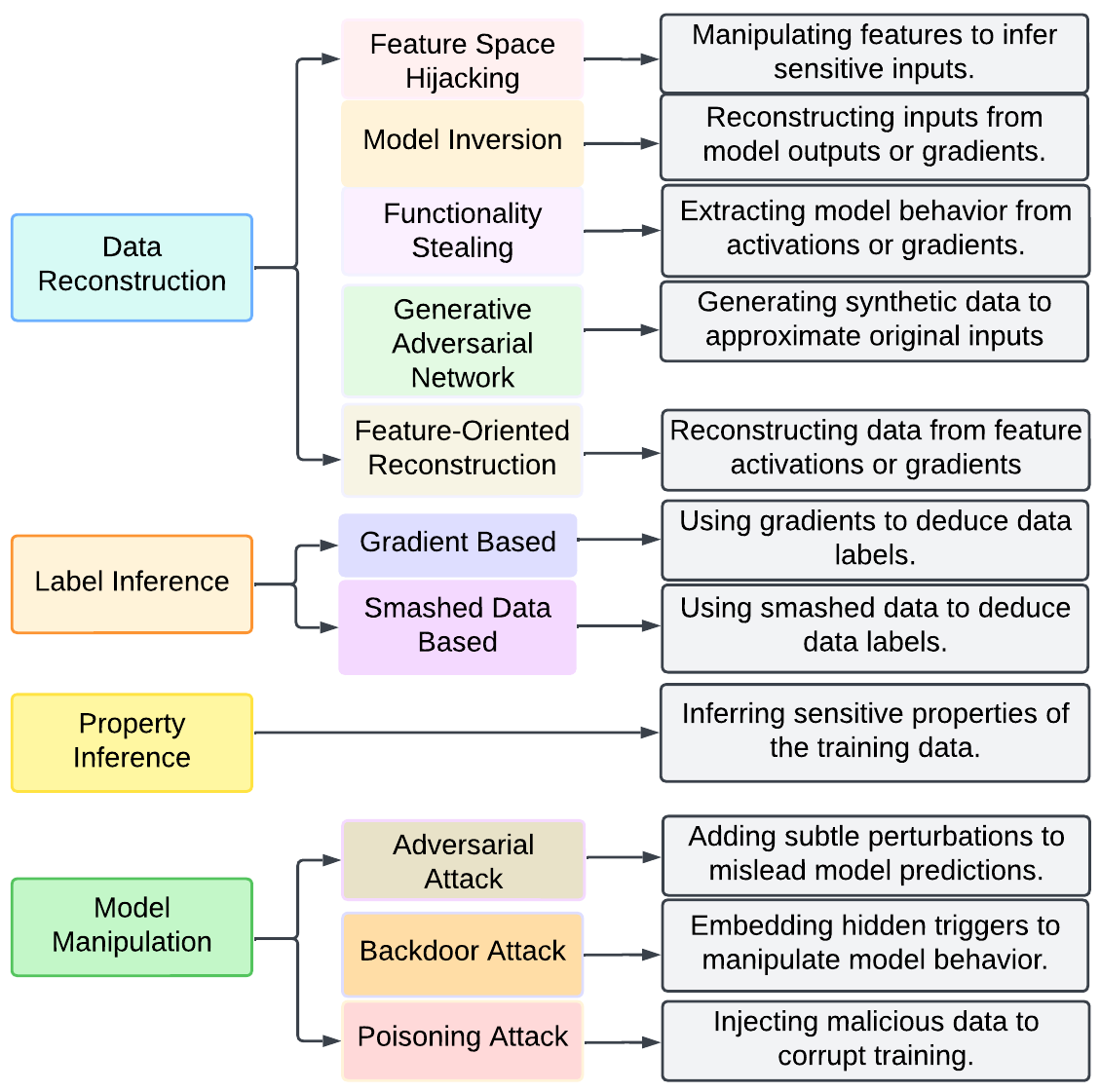} \caption{A taxonomy of attacks in SL categorizing attack vectors into: (1) Data Reconstruction, (2) Label Inference, (3) Property Inference, and (4) Model Manipulation Attacks.}
  % \Description{A taxonomy of attacks in SL categorizing attack vectors.}
  \label{fig:Attack taxonomy}
      \vspace{-1em}
\end{figure}
\subsection{Attack Strategies}
\label{sec: Attack Tecniques Employed}
\begin{figure}[htbp]
\centering
\includegraphics[width=0.28\textwidth]{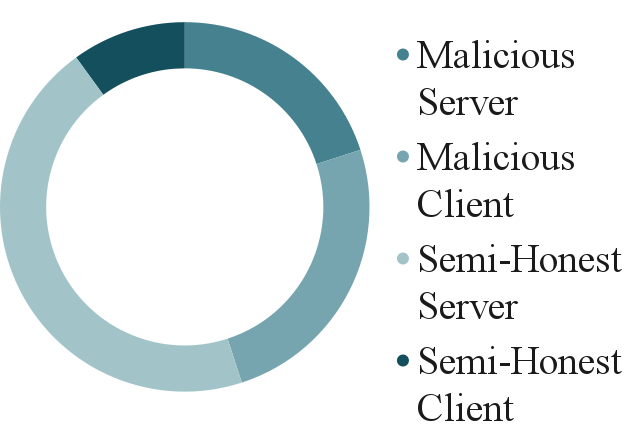}
\caption{The distribution of adversarial models of surveyed attacks in SL. }
% \Description{Graph showing the distribution of adversarial models of surveyed attack literature.}
% The figure categorizes different entities based on their adversarial behavior: Semi-Honest Servers (SHS), which dominate with the highest count, following protocols but potentially engaging in passive attacks; Malicious Clients (MC) and Malicious Servers (MS), which actively manipulate the training process to extract private information or inject adversarial data; and Semi-Honest Clients (SHC), which comply with SL protocols but may unintentionally leak information.}
\label{fig:Adversarial Model}
    \vspace{-1em}
\end{figure}

The attacks exploit specific ML principles and methodologies. In this section, we categorize them into four main strategies: data reconstruction attacks (Section~\ref{sec:reconstruction}), label inference attacks (Section~\ref{Label Inference Attacks}), property inference attacks (Section~\ref{sec:property}), and model manipulation attacks (Section~\ref{Model Manipulation Attacks}). We further refine these categories into subcategories in Figure~\ref{fig:Attack taxonomy}, which presents a structured taxonomy of attacks, systematically highlighting key attack vectors. Additionally, Figure~\ref{fig:Adversarial Model} illustrates the distribution of adversarial attack models.
% \begin{figure}
%   \centering  \includegraphics[width=0.5\textwidth]{Attack/SL Architecture.png} \caption{SL Architecture}  
%   \label{fig:SL Architecture}
% \end{figure}
% \begin{figure}
%   \centering  \includegraphics[width=0.5\textwidth]{Attack/Server Client Scenarios in Split Learning Attacks.png} \caption{Server Client Scenarios in Split Learning Attacks}  
%   \label{fig:Server Client Scenarios in Split Learning Attacks}
% \end{figure}
% \begin{figure}
%   \centering  \includegraphics[width=0.5\textwidth]{Attack/SL Types _ Server Client Scenarios.png} \caption{Taxonomy of attacks in split learning across different architectures (VanSL, USL, VSL) and client-server scenarios (SSSC, SSMC), categorized into Data Reconstruction, Label Inference, Property Inference, and Model Manipulation.}  
%   \label{fig:SL Types _ Server Client Scenarios}
%     \vspace{-1em}
% \end{figure}
% \begin{figure}
%   \centering  \includegraphics[width=0.5\textwidth]{Attack/Server Client Scenarios in Split Learning Attacks.png} \caption{Server Client Scenarios in Split Learning Attacks}  
%   \label{fig:Server Client Scenarios in Split Learning Attacks}
% \end{figure}
\subsubsection{Data Reconstruction Attacks}\label{sec:reconstruction}

Data Reconstruction Attacks represent a significant privacy vulnerability in SL systems, wherein the adversary attempts to recover sensitive training data. These attacks leverage model characteristics, including gradients, confidence scores, and internal representations, to reconstruct private training inputs. The adversary optimizes a reconstruction function to approximate the original input \( x \), formulated as:
\begin{equation}
\tilde{x} = \arg\min_{\tilde{x}^*} \mathcal{L}_{\text{rec}}(f_c(\tilde{x}^*), z_c) + \lambda R(\tilde{x}^*)
\end{equation}
where \( \tilde{x} \) is the reconstructed input, and \( \tilde{x}^* \) is iteratively refined to align with $x$. The reconstruction loss \( \mathcal{L}_{\text{rec}} \) minimizes the difference between the reconstructed and original feature representations. At the same time, regularization \( R(\tilde{x}^*) \) imposes constraints such as Total Variation (TV) loss~\cite{rudin1992nonlinear} or adversarial regularization~\cite{nasr2018machine}. The term \( \lambda \) controls the balance between accuracy and constraint enforcement. The primary methodologies are (i) Feature Space Hijacking Attack (FSHA), (ii) Model Inversion Attack, (iii) Functionality Stealing, (iv) Generative Adversarial Network (GAN), and (v) Feature Reconstruction. We further provide detailed explanations of each subtype of data reconstruction attack in Supplementary Material~\ref{sec:Details of Data Reconstruction Attacks}. 
\noindent\paragraph{\ref{sec:reconstruction}.1 Feature Space Hijacking Attack (FSHA)}\label{FSHA}
FSHA~\cite{Unleashing_tiger,gawron2022feature} is a data reconstruction attack that exploits feature representations in SL settings. By leveraging adversarial learning, a technique where models are trained using adversarial examples, intentionally crafted inputs designed to mislead the model, FSHA enables the adversary to approximate private client data from the exchanged feature space. This attack employs a three-component system: a pilot network ($\tilde{f_c}$) that defines the target feature space, an inverse network ($\tilde{f_c}^{-1}$) that reconstructs inputs from features, and a discriminator ($D$) that guides the mapping learning through adversarial training. 
\noindent\paragraph{\ref{sec:reconstruction}.2 Model Inversion:}
Model Inversion is an attack technique where an adversary attempts to reconstruct private training data by exploiting a model's parameters, outputs, or gradients. As demonstrated by~\cite{he2019model,erdougan2022unsplit}, a malicious party can leverage the information received during the SL process to recover client data. This approach involves an optimization procedure where the adversary generates synthetic inputs that produce activations or gradients matching those observed during training when fed through the model. By iteratively minimizing the distance between these synthetic outputs and the observed ones, often using mean squared error (MSE) as the objective function, the adversary can approximate the original private data. 
\noindent\paragraph{\ref{sec:reconstruction}.3 Functionality Stealing:}
Functionality stealing~\cite{zhang2024functionality} is an attack where an adversary replicates a model’s behavior. By mimicking its functionality, the attacker approximates the original model and extracts sensitive information. The adversary can train a pseudo-client model (\( \tilde{f}_c \)) that functionally mimics $f_c$ without knowing its structure. Once the functionality is stolen, the adversary can train a reverse mapping function to transform features back into $x$, effectively compromising client privacy across multiple clients without modification.
\noindent\paragraph{\ref{sec:reconstruction}.4 Generative Adversarial Network (GAN)}
\revised{
GANs have emerged as powerful tools for data reconstruction attacks~\cite{zeng2025gan,mao2023securesplit} in SL environments. This approach involves training the generator $G$ to produce synthetic inputs that yield feature representations or gradients matching those observed during training. The discriminator $D$ helps refine these reconstructions by providing feedback on their realism. GAN-based attacks are concerning because they can operate effectively with limited information and can progressively improve reconstruction quality through the adversarial training process. 
% Similarly, Zeng et al.~\cite{zeng2025gan} enables the semi-honest server in the USL setup to reconstruct private client data by leveraging GANs. The attack begins with the construction of a shadow model, wherein the server initializes an approximation of the client’s model using an auxiliary dataset. The server employs a GAN discriminator alongside cross-entropy loss to ensure the shadow model's output closely aligns with the $z_c$. Once the shadow model is trained, the server constructs an inverse model that maps the feature space back to the original input space. Using MSE loss, this inverse model is optimized to refine data reconstruction. While they exploit the vulnerabilities in server-side learning, SL remains susceptible to attacks from malicious clients when extended to multi-client SL. The client being an honest but curious adversary, extracts the server model by training an alternative model using knowledge distillation. By minimizing KL divergence loss, the client aligns the alternative model’s output with that of the global server model, effectively acquiring an approximation of the overall data distribution. After this, the client proceeds with feature space inversion, employing GAN-based optimization to infer missing class samples from other clients. Instead of reconstructing data directly in pixel space, it maps the search into a low-dimensional noise space, which reduces complexity and enhances inference efficiency.
\noindent\paragraph{\ref{sec:reconstruction}.5 Feature Reconstruction}
Feature reconstruction is an attack that aims to recover the original input data from intermediate feature representations exchanged during model training. By leveraging statistical measures~\cite{gretton2012optimal,long2015learning} and adversarial learning, attackers can approximate private data to exploit the representation preferences encoded in $z_c$~\cite{xu2024stealthy,zhu2023passive}. }
\subsubsection{Label Inference Attacks} \label{Label Inference Attacks}
Label inference attacks exploit the correlation between model updates and labels to infer client information. By analyzing shared gradients or smashed data during training, these attacks leverage inherent patterns in the learning process to reconstruct labels. They are primarily categorized into gradient-based and smashed data-based label inferences. % (see Appendix \ref{sec:labelinferenceExtended} for a detailed discussion on attack methodologies).
\noindent\paragraph{\ref{Label Inference Attacks}.1 Gradient-Based Label Inference:} These attacks exploit the correlation between gradient updates and label distributions to infer client information. We provide a detailed explanation of gradient-based label inference attacks in Supplementary Material~\ref{sec:labelinferenceExtended}. Three methods dominate the literature for performing gradient-based label inference attacks: (i) similarity-based techniques~\cite{liu2024similarity,li2021label} that compare gradients directly by selecting the expected label $({y_\text{exp}})$ and minimizing the difference between the observed gradient $(\nabla_c)$ and expected gradient $(\nabla_{exp})$:
\begin{equation}
    \tilde{y}= \arg \min_{y_\text{exp}} \| \nabla_c - \nabla_{exp} \|^2
\end{equation}
where \(\|\cdot\|\) represents the similarity metric, such as Euclidean norm (\(\ell_2\)-norm) or cosine similarity.
(ii) loss function-based inference~\cite{erdougan2022unsplit, xie2023label} that iteratively refines predictions in searching for the most expected label $y_\text{exp}$ by minimizing MSE or cross-entropy loss:
\begin{equation}
    \tilde{y}= \arg \min_{y_\text{exp}} \mathcal{L}_{\text{adv}}( \nabla_c - \nabla_{exp})\end{equation}
where $(\mathcal{L}_{\text{adv}})$ is the adversary loss function measuring the difference between $\nabla_c$ and $\nabla_{exp}$,
and (iii) surrogate model optimization~\cite{kariyappa2023exploit, zhao2024splitaum,bai2023villain}, where an adversary iteratively refines the estimated label by minimizing the loss function using gradient-based optimization. Unlike direct gradient matching or similarity-based inference, this method leverages a surrogate model to approximate the relationship between $x$, $\nabla$, and $y$, allowing for a structured reconstruction of private labels. At each iteration \( t \), the adversary updates the estimated label \( \tilde{y}{(t)} \) by performing a gradient step that minimizes the loss function:
\begin{equation}
\tilde{y}^{(t+1)} = \tilde{y}^{(t)} - \eta \frac{\partial \mathcal{L}{\text{adv}}(\nabla_c, \nabla{exp}(\tilde{y}^{(t)}))}{\partial \tilde{y}}
\end{equation}where $\eta$ is the learning rate that controls the step size for updating the label estimate.

\noindent\paragraph{\ref{Label Inference Attacks}.2 Smashed Data-Based Label Inference:} 
In smashed data-based label inference attacks, an adversary aims to reconstruct private labels by analyzing the structural and semantic properties of $z$ received by the adversary in an SL setup. Unlike gradient-based attacks, these methods do not require gradient updates but instead rely on the inherent information encoded in $z$. The adversary infers labels by minimizing a predefined loss function $(\mathcal{L}_{\text{adv}})$, which measures the difference between the observed $z$ and reference embeddings $(z_\text{exp})$. This general process is formulated as:
\begin{equation}
    \tilde{y}= \arg \min_{y_\text{exp}} \mathcal{L}_{\text{adv}}( \mathcal{F}(z), \mathcal{F}(z_\text{exp}))
\end{equation}
where, $\mathcal{F}$ represents the adversarial function. We detail the smash-based label inference attacks in Supplementary Material~\ref{Details of Smashed-Based Label Inference}. \revised{Researchers have explored three primary approaches for label inference leveraging smashed data in SL: (i) distance-based matching \cite{liu2024similarity}, which measures the similarity between observed and reference embeddings using metrics like Euclidean distance; (ii) clustering-based inference \cite{liu2024similarity, zhu2023passive}, which groups similar embeddings to identify patterns in labels; and (iii) transfer learning-based inference \cite{liu2024similarity, huang2023pixel}, which leverages pre-trained models or generators to extract label information from embeddings. These techniques demonstrate that even without gradient information adversaries can successfully infer private labels by exploiting the semantic properties preserved in the shared intermediate representations.}

% Liu et al.~\cite{liu2024similarity} introduced three primary approaches for label inference using smashed data: Euclidean Distance-Based Matching, Clustering-Based Inference, and Transfer Learning-Based Inference, demonstrating that even without gradients, label inference remains possible through semantic similarities in $z_c$. Zhu et al.~\cite{zhu2023passive} extended label inference to USL, where the client retains both the first and last model layers. They introduced a label simulator $\tilde{h}$ to approximate the client's final layers and infer labels, mitigating overfitting through random label flipping. In contrast, Huang et al.~\cite{huang2023pixel} targeted SFL with a two-phase attack that reconstructs pixel-wise accurate training data. Their method employs a pseudo-sample generator for class-balanced examples and cycle-consistency losses to refine image reconstruction. While Liu et al.~\cite{liu2024similarity} and Zhu et al.~\cite{zhu2023passive} focus on label inference, Huang et al.~\cite{huang2023pixel} shift toward reconstructing $x$, highlighting different vulnerabilities in SL frameworks. 
\subsubsection{Property Inference Attacks}\label{sec:property}
Property inference attacks extract sensitive attributes or statistical patterns from client data without requiring complete data reconstruction. Unlike reconstruction attacks, these approaches focus on inferring specific properties such as demographic information, class distributions, or other sensitive characteristics embedded within the private dataset. As demonstrated in~\cite{Unleashing_tiger, mao2023securesplit}, property inference attacks in SL typically leverage the intermediate representations or model updates shared during the collaborative training process. The approach involves an adversary creating a specialized classifier $(C)$ that maps the observed smashed data or gradients to property labels. By analyzing patterns in the information, an adversary can infer properties that were never intended to be shared. The attack can be formulated as:
\begin{equation}
A_{PIA} = \arg\min_F \sum_{x_i \in X} \mathcal{L}_{adv}(F(T(x_i)), l_i), \quad l_i \in {0,1}
\end{equation}
where $( F )$ represents the adversarial inference model, $( T )$ is the function that processes client data and produces the observable artifacts, and $( l_i )$ indicates the property being inferred. These attacks are particularly concerning because they can reveal sensitive information while the participating entities believe they are only sharing task-relevant features.
\subsubsection{Model Manipulation Attacks} \label{Model Manipulation Attacks} This section explores attacks that compromise SL model integrity, categorizing them into: (i) adversarial attacks that perturb inputs to cause misclassification, and (ii) backdoor and poisoning attacks that manipulate training data to induce malicious behavior or degrade performance (see Supplementary Material~\ref{sec:Model ManipulationExtended}).
\paragraph{\ref{Model Manipulation Attacks}.1 Adversarial Attacks.}
Adversarial attacks manipulate intermediate feature representations by introducing perturbations that induce misclassification. These attacks can be categorized as non-targeted attacks~\cite{fan2023robustness} and targeted attacks~\cite{he2024advusl}. We provide a summary of adversarial attacks and detail them in Supplementary Material~\ref{sec:Adversarial Attacks Extended}. The objective of a non-targeted attack is to maximize the difference between the perturbed representation and the clean feature representation \( z \). This is achieved by often using cosine similarity or Euclidean distance:
\begin{equation}\label{untargeted}
\xi^* = \arg\max_{\|\xi\|_{\infty} \leq \epsilon} \mathcal{L}_{\text{attack}}(z, \xi + z)
\end{equation}
where \( \xi^* \) is the optimal adversarial perturbation, \( \xi \) is the adversarial perturbation applied to \( z \), and \( \|\xi\|_{\infty} \leq \epsilon \) ensures that the perturbation remains within the allowed bound. However, in a targeted attack, the adversary modifies \( z \) such that they move toward a predefined target embedding \( z_t \), forcing the model to produce a specific incorrect prediction. This is achieved by minimizing (rather than maximizing) the attack loss function in Equation~\ref{untargeted}. % but with the objective changed from maximization to minimization. 
The reason %for the change in the objective function 
is that in a non-targeted attack, the adversary aims to \textit{maximize} the deviation of the perturbed representation from the original feature representation, making the output unpredictable and unreliable. In contrast, a targeted attack seeks to \textit{minimize} the difference between the perturbed representation and a specific target representation \( z_t \), effectively steering the model toward a controlled misclassifications. 

\paragraph{\ref{Model Manipulation Attacks}.2 Backdoor and Poisoning Attacks.}\label{Backdoor and Poisoning Attack}
SL models are vulnerable to adversarial manipulations that exploit weaknesses in the training process, particularly through poisoning attacks and backdoor attacks. \textit{Poisoning attacks} manipulate the training process by modifying the dataset to degrade model performance. Let \( x, y \) denote the original training data and labels, and \( \mathcal{A}(x, y) \) be the adversarial poisoning function that generates a poisoned dataset \( x', y' \). The model, parameterized by \( \theta \), updates its parameters to \( \theta^* \) after training on poisoned data, leading to incorrect decision boundaries. The adversary’s objective is to maximize classification errors, formulated as a minimization problem where the model, trained on poisoned data, unknowingly optimizes its loss function in a way that degrades generalization and increases misclassification:
\begin{equation}
\theta^* = \arg\min_{\theta} \mathbb{E}_{(x', y') \sim \mathcal{A}(x, y)} \mathcal{L}(f_{\theta}(x'), y')
\end{equation}
where \( f_{\theta}(x') \) is the model's output, and \( \mathcal{L}(f_{\theta}(X'), Y') \) measures the classification loss on the poisoned dataset. The expected loss function \( \mathbb{E}_{(x', y') \sim \mathcal{A}(x, y)} \) ensures the poisoning effect generalizes across training samples.
\textit{Backdoor attacks} implant hidden triggers into the training process, enabling adversaries to manipulate model predictions when the trigger is present while maintaining normal behavior otherwise. Let \( x, y \) denote the clean training data and labels, and let \( x_b \) be the backdoor-inserted inputs with the attacker-defined target labels \( y_t \). The adversary optimizes a dual-objective function that ensures normal classification on clean samples while inducing misclassification on backdoor samples:
\begin{equation*}
   \theta^* = \arg\min_{\theta} \mathbb{E}_{(x, y) \sim \mathcal{D}_{\text{clean}}} \mathcal{L}(f_{\theta}(x), y) + \lambda \mathbb{E}_{(x_b, y_t) \sim \mathcal{D}_{\text{b}}} \mathcal{L}(f_{\theta}(x_b), y_t)  
\end{equation*}
where \( \mathcal{D}_{\text{clean}} \) represents the clean dataset used for normal training, while \( \mathcal{D}_{\text{b}} \) contains samples modified with a trigger. The adversary applies a backdoor function to \( x \), producing \( x_b \) that, when processed by the model, leads to a specific misclassification. The attacker assigns target labels \( y_t \) to these backdoor-embedded samples, ensuring that normal inputs retain their correct classification while triggered inputs are classified incorrectly. \( \lambda \) is a regularization parameter that balances the model’s ability to maintain the accuracy on clean inputs while embedding the backdoor functionality (see Supplementary Material~\ref{sec:Backdoor and Poisoning Attack Extended} for details of backdoor and poisoning attacks.)
\par\noindent\textbf{a. Label Flipping Attacks:}
%\label{sec:Backdoor and Poisoning Attack Extended}
\revised{Label flipping attacks represent a fundamental poisoning strategy in SL, where adversaries deliberately alter class labels within the training data~\cite{kohankhaki2023detecting, gajbhiye2022data, ismail2023analyzing}. These attacks are executed by malicious clients who manipulate their local datasets before participating in the collaborative training process. The general approach involves systematically modifying the true labels of training samples to incorrect ones, either through targeted flipping (changing specific source classes to pre-selected target classes), untargeted flipping (random mislabeling across multiple classes), or distance-based optimization (selecting target classes that maximize classification errors based on feature proximity). The effectiveness of label flipping attacks increases with the poisoning rate. For instance, even 10\% malicious clients reduced class recall by 12.58\% on CIFAR-10 and up to 47.31\% with 50\% malicious clients~\cite{gajbhiye2022data}. Similar trends were observed by Ismail and Shukla~\cite{ismail2023analyzing}, where untargeted and distance-based attacks caused greater accuracy degradation than targeted attacks in both MNIST and ECG datasets, confirming their higher impact across domains.}
% Label-flipping attacks are a type of data poisoning attack where adversaries intentionally modify the class labels of training samples to mislead the model and degrade its classification accuracy. Kohankhaki et al.~\cite{kohankhaki2023detecting} investigated static label flipping attacks in VanSL, where malicious clients systematically modify class labels. They analyzed different poisoning scenarios by adjusting the number of malicious clients and poisoning rates, specifically manipulating ECG readings by flipping labels between normal and abnormal classes. Gajbhiye et al.~\cite{gajbhiye2022data} examined label-flipping attacks on SFL where adversaries control only their local training data. They contrasted targeted attacks (systematically mislabeling a source class as a specific target class) with untargeted attacks (random mislabeling across multiple classes). Ismail et al.~\cite{ismail2023analyzing} introduced distance-based poisoning as a third strategy alongside targeted and untargeted poisoning methods. Their approach uniquely optimizes label flipping by selecting target classes based on Euclidean distance to maximize classification errors. 

\par\noindent\textbf{b. Embedding Poisoning Attacks:}
\revised{Embedding poisoning attacks target the smashed data, manipulating the embedding space rather than raw data or labels~\cite{bai2023villain, wu2024evaluating}. The general approach involves introducing perturbations to the feature representations ($\tilde{z_i}$) before they are transmitted to the server. These perturbations are designed to be subtle enough to avoid detection while causing the model to learn unintended patterns or vulnerabilities. These attacks are effective because they directly compromise the information exchange that is fundamental to the collaborative learning process.}
% Embedding poisoning attacks manipulate feature representations within the model’s embedding space, introducing adversarial biases that degrade model performance or create hidden vulnerabilities. Bai et al.~\cite{bai2023villain} developed a two-stage attack methodology combining label inference with embedding poisoning. Following successful label inference (Section~\ref{Label Inference Attacks}), attackers poison embeddings by introducing a stealthy trigger vector \( E \), defined as $E = M \otimes (\beta \cdot \Delta)$, where \( M \) selects key embedding dimensions, \( \beta \) controls perturbation magnitude, and \( \Delta \) introduces structured variations through alternating positive and negative values. The final poisoned embedding, computed as $\tilde{z_i} = z_i \oplus E$, maintains attack stealth while implementing an exploitable backdoor. Wu et al.~\cite{wu2024evaluating} systematically evaluate SFL security against poisoning attacks with malicious clients as an adversary: \textit{dataset poisoning} modifies client-level training data; \textit{weight poisoning} manipulates model parameters; \textit{label poisoning} alters class labels for targeted misclassifications; \textit{smash poisoning} modifies intermediate data before server transmission; and \textit{hybrid approaches} combine these methods to amplify adversarial effects.

\par\noindent\textbf{c. Client-Side Backdoor Attacks:}
\revised{
Client-side backdoor attacks in SL involve malicious clients who inject hidden triggers into their local training data or model components. These attacks follow a pattern where the adversary modifies a subset of their training data to include specific trigger patterns associated with targeted misclassifications~\cite{yu2023backdoor, safeSplit2025rieger}. The approach involves training the local model component to respond to these triggers while maintaining normal performance on clean data. This can be achieved through feature manipulation, label modification, or the use of auxiliary models to distinguish between clean and backdoored samples. The distributed nature of the system makes it difficult to detect malicious behavior, allowing backdoors to persist across multiple training rounds while maintaining model utility for the primary task.}
% These attacks involve injecting hidden triggers into the local training data or model updates of a client, allowing the adversary to manipulate model predictions when the trigger is present. Yu et al.~\cite{yu2023backdoor} proposed client-side backdoor attacks, where a malicious client in VanSL injects backdoor samples by modifying features or labels. They introduced an auxiliary model to distinguish between clean and backdoor samples, enhancing attack persistence while maintaining model performance on primary tasks. Furthermore, Rieger et al.~\cite{safeSplit2025rieger} examined client-side backdoor attacks in a USL setup, where the attack process begins with trigger insertion into a subset of training data. 
\par\noindent\textbf{d. Server-Side Backdoor Attacks:}
\revised{
In server-side backdoor attacks, a malicious server compromises the integrity of the SL model. As investigated in~\cite{tajalli2023feasibility, yu2023backdoor, yu2024chronic}, these attacks leverage the server's privileged position in the training process to implant backdoor functionalities without direct access to client data. The methodology involves manipulating the shared model components or gradient updates to create hidden vulnerabilities that can be exploited later. This can be achieved through surrogate model training, feature space manipulation, or strategic modification of model updates.} 
% Tajalli et al.~\cite{tajalli2023feasibility} investigated server-side backdoor attacks in VanSL, proposing two strategies: Surrogate Client Attack and Injector Autoencoder Attack. While they suggested that VanSL is resilient to backdoor attacks, Yu et al.~\cite{yu2023backdoor} demonstrated that a malicious server can implant backdoors via feature space hijacking, aligning the client model's optimization with a shadow model. Furthermore, Yu et al.~\cite{yu2024chronic} introduced the SFI (Steal, Finetune, and Implant) attack framework, a sophisticated server-side backdoor attack in VanSL. The three-stage framework manipulates gradient updates without requiring access to raw client data.

\subsection{Attack Constraints}
% \begin{table*}[h!]
% \centering
% \renewcommand{\arraystretch}{1.5} % Adjust row height for readability
% \begin{adjustbox}{width=\textwidth}
% \begin{tabular}{|c|c|c|c|c|l|}
% \hline
% \rowcolor[HTML]{EAEAEA} 
% \textbf{Attack Constraints} & \textbf{Data Reconstruction(\(A_1\))} & \textbf{Label Inference(\(A_2\))} & \textbf{Property Inference(\(A_3\))} & \textbf{Model Manipulation (\(A_4\))} & \textbf{References} \\ \hline
% Dataset Dependency & \ding{108} & \ding{108} & \ding{109} & \ding{108}  &\cite{Unleashing_tiger,zhang2024functionality,xu2024stealthy,yu2024chronic,yu2023backdoor,he2024advusl,zhu2023passive,zeng2025gan,huang2023pixel,ismail2023analyzing}\\ \hline
% Knowledge Dependency  & \ding{108} & \ding{108} & \ding{108} & \ding{109}  &\cite{Unleashing_tiger,zhang2024functionality,fan2023robustness}\\ \hline
% Architecture Exposure & \ding{108} & \ding{108} & \ding{108} & \ding{108}  &\cite{erdougan2022unsplit,tajalli2023feasibility,he2024advusl,zhu2023passive,huang2023pixel}\\ \hline
% Label and Classification Dependency & \ding{109} & \ding{108} & \ding{109} & \ding{109}  &\cite{erdougan2022unsplit,li2021label,liu2024similarity,bai2023villain,tajalli2023feasibility}\\ \hline
% \end{tabular}
% \end{adjustbox}
% \caption{Constraints required for the successful execution of various attacks in split learning, grouped by main attack categories. \ding{108} indicates a strong requirement.}
% \label{tab:Attack_constraints}
% \vspace{-1em}
% \end{table*}
\label{sec: Attack Constraints}
Attack constraints define the technical and operational prerequisites for adversaries to execute attacks in SL successfully. %These constraints vary by attack strategies and influence the feasibility and effectiveness of the attack. 
% Table~\ref{tab:Attack_constraints} provides a comprehensive overview of the constraints for each type of attack. 
\paragraph{Dataset Constraints}
\revised{Dataset constraints influence the vulnerability landscape of SL systems. One fundamental constraint is the adversary’s ability to access domain-similar datasets, which substantially enhances the feasibility of various attack vectors~\cite{Unleashing_tiger}. Studies have shown that an exact match between the adversary’s dataset and the target data is not necessary; it is sufficient for the datasets to share similar distributional characteristics to mount effective attacks~\cite{xu2024stealthy,he2024advusl,zhu2023passive,huang2023pixel}. Furthermore, access to publicly available datasets enables the development of more sophisticated threats. For instance, adversaries can train surrogate models to reconstruct private data~\cite{zhang2024functionality,zeng2025gan} or employ shadow models to conduct backdoor insertion and poisoning attacks~\cite{yu2024chronic,yu2023backdoor,ismail2023analyzing}.%These approaches ultimately allow adversaries to manipulate model behavior and compromise system security.
}

\paragraph{Knowledge-Based Constraints}
\revised{
The assumption that the server possesses complete knowledge of the learning task significantly amplifies the effectiveness of adversarial strategies. Under this constraint, adversaries are able to design task-specific queries that exploit the model’s learning objective, thereby increasing the risk of privacy breaches~\cite{Unleashing_tiger,zhang2024functionality}. For instance, in a setting where the learning task involves medical imaging, adversaries can synthesize patient-specific scans to infer sensitive attributes. Additional assumptions, such as the server and client sharing the same optimizer, further facilitate attack success by enabling more accurate reconstruction of training data or inference of model parameters~\cite{zhang2024functionality}. %In contrast to models assuming full task knowledge, s
Some works explore scenarios where the adversary operates with only partial information about the learning objective~\cite{fan2023robustness}. %, thereby limiting the strength of the attacks.
}
\paragraph{Architecture Exposure Constraints}
Architecture exposure constraints play a crucial role in determining the vulnerability of SL systems to various attacks. The exposure to client-side architecture significantly enhances the efficacy of model manipulation~\cite{he2024advusl,huang2023pixel}, inversion~\cite{erdougan2022unsplit}, and feature reconstruction attacks~\cite{zhu2023passive}. Further expanding on this constraint, more comprehensive threat models have been proposed, where adversaries are assumed to have full knowledge of both the client's subnetwork architecture and the dataset distribution~\cite{tajalli2023feasibility}. This enhanced architectural exposure notably strengthens the potential for backdoor attacks, enabling adversaries to manipulate the model's internal structure and implement covert functionalities while evading detection.
\paragraph{Label and Classification Constraints}
%Label and classification constraints substantially impact the feasibility of inference attacks in split learning. 
Knowledge of the number of discrete labels in the dataset allows adversaries to refine their predictions and narrow the search space for classification inference~\cite{erdougan2022unsplit}. In some settings, adversaries are further assumed to employ binary classifiers to predict labels from observed model activations, a technique commonly leveraged in membership inference attacks to determine the presence of specific samples in the training set~\cite{li2021label}. Additional assumptions, such as requiring only a single labeled sample per class or having access to all participant labels, further strengthen the adversary’s capability to infer broader label information from model activations~\cite{liu2024similarity,bai2023villain}.

\subsection{Attack Effectiveness}
\label{sec: Attack Effectiveness}
%Attack effectiveness in SL varies, exposing vulnerabilities for privacy and integrity. Below, we provide the key influencing factors:
% This section synthesizes the insights presented in Table~\ref{tab:attack_effectiveness_with_references}.
\paragraph{Domain-Independent Attacks}
Such attacks do not rely on specific datasets, architectures, parameter distribution, or optimization methods. Unlike traditional attacks that exploit domain-specific patterns, these attacks leverage fundamental weaknesses in the SL process, allowing them to generalize across multiple settings. For instance,~\cite{Unleashing_tiger,zhang2024functionality,xu2024stealthy,liu2024similarity,li2021label,kariyappa2023exploit,bai2023villain,erdougan2022unsplit} attacks have been shown to reconstruct input data and inference labels across diverse learning environments, demonstrating that their effectiveness is not constrained to a particular application. 
\paragraph{Stealth}
%Another major factor contributing to the effectiveness of SL attacks is their stealth, which allows them to operate undetected. 
In traditional security frameworks, attacks are often identified by monitoring anomalies in system behavior. However, many SL attacks~\cite{Unleashing_tiger,zhang2024functionality,erdougan2022unsplit,xu2024stealthy,liu2024similarity,yu2024chronic,he2024advusl,zhu2023passive,zhao2024splitaum,safeSplit2025rieger,xie2023label,gajbhiye2022data,ismail2023analyzing} evade detection by ensuring that they do not introduce observable deviations in gradient updates or model performance. For instance, distance correlation minimization is leveraged in~\cite{Unleashing_tiger,erdougan2022unsplit,xu2024stealthy} to launch stealthy attacks, where gradient updates are subtly manipulated while preserving the overall model behavior. 
\paragraph{Defeating Differential Privacy}
Differential Privacy (DP)~\cite{dwork2006differential} is recognized as one of the most effective privacy-preserving methods in machine learning. It operates by introducing controlled random noise into training updates, thereby ensuring that the inclusion or exclusion of any individual data point does not significantly alter the model’s output (detailed in Section~\ref{sec:DataPerturbation}). This mechanism provides theoretical guarantees that adversaries cannot precisely infer the presence of a specific data sample. However, recent studies~\cite{gawron2022feature,zhang2024functionality,li2021label,liu2024similarity,bai2023villain,zhao2024splitaum,huang2023pixel} have demonstrated that DP alone is insufficient in defending against advanced SL attacks. The primary limitation of DP in the SL setting is that it focuses on protecting individual data contributions rather than preventing full-scale data reconstruction. While DP effectively limits the sensitivity of activations and gradient updates, adversaries can still exploit structural and statistical properties of gradients to recover private information at a more aggregated level. Moreover, adversaries employing distance correlation minimization techniques~\cite{Unleashing_tiger,erdougan2022unsplit,xu2024stealthy} have been shown to circumvent DP protections. By iteratively optimizing gradient representations, adversaries can reconstruct highly accurate approximations of the original input data, reducing the effectiveness of DP.
\paragraph{SL Variants}
The effectiveness of attacks in SL is not confined to a single variant but extends across all architectures~\cite{zhang2024functionality,fan2023robustness,zhu2023passive,zhao2024splitaum}, demonstrating a fundamental vulnerability. Since all SL variants inherently rely on exchanging intermediate activations or gradients between the clients and the servers, adversaries can exploit this communication to infer sensitive information, manipulate learning dynamics, or introduce adversarial perturbations. The structural modifications between VanSL, USL, and Hybrid SL do not provide sufficient protection, as the core vulnerability (gradient leakage and model influence) remains consistent across variants.

%% file: defensemechanisms.tex
\section{Defense For Split Learning}
\label{sec:Defense Mechanisms}

In Section~\ref{sec: Threats and Attack Scenarios}, we explored how adversaries exploit SL vulnerabilities. In response, various defense strategies have emerged. This section systematically analyzes defense mechanisms based on \emph{defense strategies} (Section~\ref{sec:DefenseTechniquesEmployed}), \emph{operational constraints} (Section~\ref{sec:DefenseConstraints}), and \emph{effectiveness} (Section~\ref{sec:DefenseEffectiveness}), with a summary in Table~\ref{tab:general table}, Supplementary Material~\ref{sec:supplementaryTable}. 
\begin{figure}[htbp] 
  \centering  \includegraphics[width=0.5\textwidth]{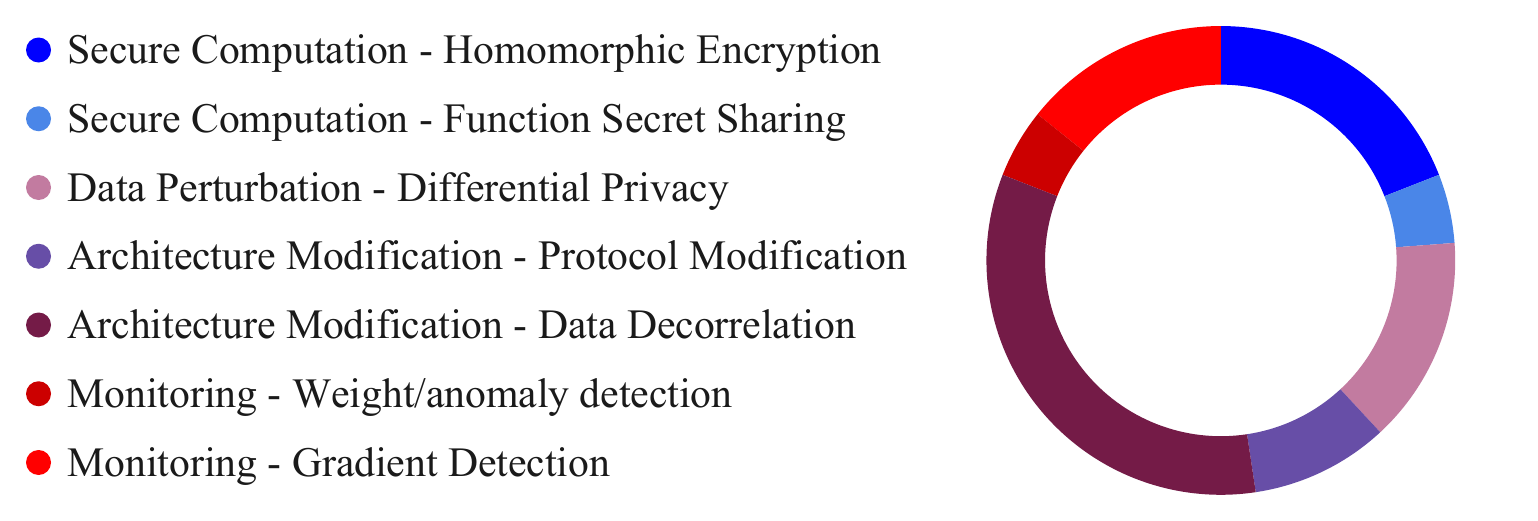} \caption{The distribution of defense mechanisms classified in the surveyed literature.}  
  % \Description{Figure showing distribution of defense mechanisms classified in the surveyed literature.}
  \label{fig:Defense mechanisms considered in papers Pie Defense}
      \vspace{-1em}
\end{figure}

\subsection{Defense Strategies} \label{sec:DefenseTechniquesEmployed}
\begin{figure}[htbp] 
    \centering
    \includegraphics[width=0.9\linewidth]{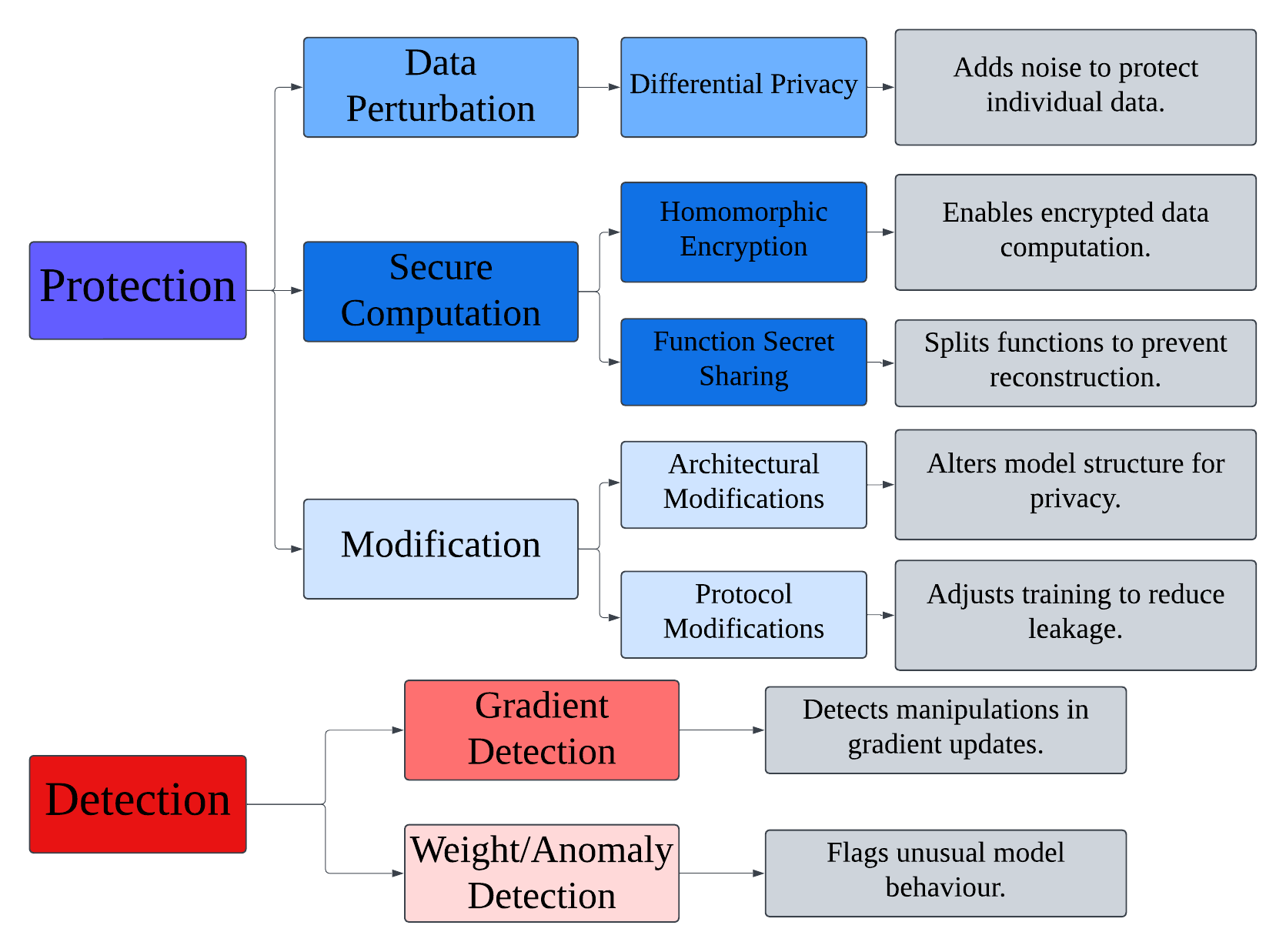}
    \caption{Taxonomy of Defense Strategies}
    % \Description{A diagram that categorizes different defense strategies.}
    \label{fig:defense-taxonomy}
    \vspace{-1em}
\end{figure}

In this section, we present defense strategies, first categorizing them into protection and detection mechanisms, followed by their subcategories. The distribution and taxonomy of these strategies in the surveyed literature is illustrated in Figures~\ref{fig:Defense mechanisms considered in papers Pie Defense}, and \ref{fig:defense-taxonomy}, respectively.

% In this section, we classify defense techniques based on their primary function in securing SL against adversarial threats. We categorize them into Protection Mechanisms (Section~\ref{sec:ProtectionMechanisms}), which proactively prevent information leakage and manipulation, and Detection Mechanisms (Section~\ref{sec:DetectionMechanisms}), which focus on identifying and mitigating adversarial actions by terminating the protocol or excluding compromised participants. Each subsection further explores specific methods within these categories, detailing their core principles and the threats they address. 

% Within each category, we further distinguish subcategories (e.g., data perturbation, secure computation, architecture-level modifications) based on the \emph{techniques} employed and the \emph{threat models} they mitigate. As in Section~\ref{sec: Threats and Attack Scenarios}, we note that each defense strategy often assumes either semi-honest or malicious adversaries and must contend with practical considerations such as accuracy trade-offs, computational overhead, and implementation constraints. This classification can be observed in Figure~\ref{fig:defense-taxonomy}.

\subsubsection{Protection Mechanisms}
\label{sec:ProtectionMechanisms}
Protection mechanisms aim to secure parties preemptively before attacks occur. These approaches generally conceal or limit the information an adversary can extract, often assuming a \emph{semi-honest} threat model. %The three main subfields of protection mechanisms are \emph{data perturbation}, \emph{secure computation}, and \emph{architecture-level modifications}.

\paragraph{\ref{sec:ProtectionMechanisms}.1 Data Perturbation:}
\label{sec:DataPerturbation}In this approach, clients intentionally \emph{modify} (e.g., add noise to) data or intermediate representations, aiming to obscure sensitive information while preserving sufficient utility for training. \emph{Differential privacy} (DP) gained prominence due to its well-defined mathematical framework that achieves provable formal privacy guarantees. By calibrating noise according to \((\varepsilon,\delta)\)-DP, data contributors can formally bound the risk. The privacy guarantee is established through the definition in Equation~\eqref{eq:DP}, ensuring that even if an adversary observes these shared activations, they cannot confidently distinguish whether a specific data point was included in the dataset \( \mathcal{X} \).
% \begin{definition}[$(\varepsilon, \delta)$-Differential Privacy]
% A randomized mechanism \( \mathcal{M} \) satisfies \((\varepsilon, \delta)\)-differential privacy if for all neighboring datasets \( \mathcal{X} \) and \( \mathcal{X}' \) differing in at most one entry, and for all measurable subsets \( S \) of the output space:
\begin{equation}
\Pr[\mathcal{M}(\mathcal{X}) \in S] \leq e^{\varepsilon} \Pr[\mathcal{M}(\mathcal{X}') \in S] + \delta.
\label{eq:DP}
\end{equation}
where $\mathcal{M}$ is a randomized mechanism  that takes a dataset $\mathcal{X}$ as input and $\mathcal{M}$ satisfies $(\varepsilon,\delta)$-differential privacy for all neighboring datasets \( \mathcal{X} \) and \( \mathcal{X}' \) differing in at most one entry. Here, $\varepsilon$ defines the privacy budget and $\delta$ is a small probability of violating strict $\varepsilon$-DP. To achieve differential privacy, noise from distributions such as Laplace, Gaussian, or uniform is often added to the data or query results to obscure the influence of any single individual. In the context of SL, DP is often applied to perturb the activations exchanged between the client and server, preventing sensitive information leakage through intermediate representations. Nonetheless, DP methods often entail a trade-off between privacy and model accuracy. %Even with DP, adversaries are able to achieve success under noisy environments employed by these techniques. 
Key challenges in these schemes include:
\begin{itemize}[nosep,leftmargin=*]
    \item \emph{Accuracy vs. Noise}: Elevated noise degrades model performance.
    \item \emph{Scalability}: Tuning DP parameters for high-dimensional data or large-scale models is non-trivial.
\end{itemize}
Several papers \cite{PrivateMail,pham2024e,gawron2022feature} investigate how DP needs to be applied in terms of both achieving privacy and accuracy in various SL setups. %As the main struggle of research in DP is not to maintain privacy but achieving privacy in an efficient and accurate manner for SL, these papers discuss not DP specific techniques but proper application of DP.

\paragraph{\ref{sec:ProtectionMechanisms}.2 Secure Computation:}
\label{sec:SecureComputation}
Cryptographic methods offer formal security guarantees by preventing adversaries from accessing raw data, with each cryptographic scheme preserving different properties. \emph{Homomorphic encryption (HE)} enables computations on encrypted data. Formally, if \( E \) represents the encryption function, \( D \) the decryption function, and \( \circ \) a supported operation (such as addition or multiplication) under HE, then an HE scheme satisfies the property:
\[
D(E(x) \circ E(y)) = x \circ y
\]
where \( x \) and \( y \) are plaintext values, and \( E(x) \) and \( E(y) \) are their corresponding ciphertexts. Each encryption scheme provides different functionalities, and in the context of SL, fully homomorphic encryption is widely preferred due to its ability to support arbitrary polynomial-time computations on encrypted data without requiring decryption.

% as illustrated in Equation~\ref{eq:Homomorphic Encryption}, which states that a circuit \( C \) consisting of summation and multiplication operations on vectors can be equivalently applied to ciphertexts corresponding to plaintexts using an evaluation function. In other words, the server can perform machine learning operations without knowing the data or the model weights.

% \begin{equation}
% \text{Eval}(C, \mathbf{ct}_1, \dots, \mathbf{ct}_n) = \mathbf{ct}_{\text{out}}; \text{Dec}(\mathbf{ct}_{\text{out}}) = C(\mathbf{pt}_1, \dots, \mathbf{pt}_n)
% \label{eq:Homomorphic Encryption}
% \end{equation}

% Here in this equation $\text{Dec}(\mathbf{ct}_i) = \mathbf{pt}_i$

On the other hand, \emph{function secret sharing (FSS)} is a cryptographic technique that allows a function to be split into $n$ shares distributed among parties, such that the original function can be reconstructed by combining the shares. Below, we present the formal definition of FSS as references in~\cite{boyle2015function}.

For \( n \in \mathbb{N} \), an \( n \)-party FSS scheme with respect to a share output decoder \( \text{DEC} = (S_1, \dots, S_n, R, \text{Dec}) \) and a function class \( \mathcal{F} \) is a pair of probabilistic polynomial-time algorithms \( (\text{Gen}, \text{Eval}) \) defined as:

\begin{itemize}
    \item \textbf{Key Generation:}  
    The algorithm \( \text{Gen}(1^\lambda, f) \) with a security parameter \( \lambda \) and a function description \( f \in \mathcal{F} \), outputs \( n \) secret keys \( (k_1, \dots, k_n) \).
    
    \item \textbf{Evaluation:}  
    Each party \( i \) runs \( \text{Eval}(i, k_i, x) \), with a party index \( i \), the corresponding key \( k_i \), and an input \( x \). It outputs a value \( y_i \in S_i \), representing the party's share of \( f(x) \).
\end{itemize}

For example, in a setting with two semi-honest servers that assumed to not collude, neither server can reconstruct the training function independently, even though together they can compute the intended final result, as  $f(x) = f_1(x) \oplus f_2(x)$ where $f_1, f_2$ are computationally indistinguishable.

% In a split learning setup as in \cite{khan2024make} with two non-colluding, semi-honest servers $S_1$ and $S_2$, the training function $f$ is split into shares $f_1$ and $f_2$, ensuring neither server can reconstruct $f$ independently. Given an input $x$, each server computes $y_1 = f_1(x)$ and $y_2 = f_2(x)$, with the final result obtained as $f(x) = y_1 \oplus y_2$. Since $f_1$ and $f_2$ are computationally indistinguishable, no individual server learns the original function or input, preserving privacy. 

%In the SL context, HE is the predominant technique, for enabling computation on encrypted data without decryption. %Advancements in Fully Homomorphic Encryption (FHE) and CKKS-based schemes~\cite{CKKS_Homomorphic} support essential operations for neural network training. 

% In the context of SL, HE is the predominant technique, as it enables computations on encrypted data without decryption. With the development of FHE and CKKS-based \cite{CKKS_Homomorphic} schemes which support the homomorphic addition and multiplication operations essential for neural network training. These approaches maintain model accuracy while providing robust security; however, their computational overhead may render them impractical for resource-constrained environments. 

% Earlier work in this category has explored various cryptographic primitives such as Function Secret Sharing (FSS) and oblivious transfer protocols.
The primary objective here is not merely to ensure data privacy, since encryption with well-established cryptographic schemes already provides a rigorous security foundation, but to optimize the application of these strategies for efficiency and practicality, particularly in resource-constrained environments. Khann et al. \cite{khan2024make} explore a two-server FSS approach, examining how different secure computation techniques can be adapted for SL.

Other works~\cite{CURE2024Kanpak, splitHE, SplitWithoutALeak, LoveHate} integrate HE into SL leveraging both cryptographic techniques and adversarial training methodologies to enhance security while maintaining computational feasibility. 
While these methods maintain model accuracy and provide robust security, their computational overhead can be impractical for resource-constrained environments as discussed in~\cite{CURE2024Kanpak}, for computational resource-restricted environments additional considerations need to be done to manage the training in a feasible way.

\paragraph{\ref{sec:ProtectionMechanisms}.3 Structural and Procedural Modifications:}
\label{sec:StructuralEnhancements}
Defense strategies in this section modify the model architecture or the protocol to limit information leakage and maintain privacy. We provide the details of two subcategories (architectural modifications and protocol modifications) below: 

\par\noindent\textbf{a. Architectural Modifications:}
\label{sec:ArchitecturalModifications}
Architectural modifications introduce structural changes to SL to prevent reconstruction attacks. Abuadbba et al.~\cite{Abuadbba2020} investigate the effects of shrinking the cut layer on the effectiveness of FSHA. Building on this, Pham et al.~\cite{pham2022binarizing} introduce a binarized neural network approach combined with DP to analyze its impact on both security and efficiency. Another important approach leverages \emph{specialized activation functions}, such as the one presented in Equation~\eqref{eq:R3eLU} by Mao et al.~\cite{mao2023securesplit}, which strategically introduce randomness into activation outputs to obscure patterns and minimize the information leakage to potential adversaries.
\begin{equation}
\text{R}^3\text{eLU}(v) =
\begin{cases}
\max(0, v + z), & \text{with probability } p, \\
0, & \text{with probability } (1 - p).
\end{cases}
\label{eq:R3eLU}
\end{equation}
Additionally, mechanisms designed to de-correlate transmitted data have been proposed in~\cite{mao2023securesplit, DISCO, PSLF2023Wan, turina2021fsl, khowaja2024foesfooled}, demonstrating various techniques to mitigate information leakage while preserving utility.

\par\noindent\textbf{b. Protocol Modifications:}
\label{sec:ProtocolModifications}
Beyond structural adjustments, refining the protocol that governs data exchange in SL can substantially strengthen privacy. A notable example is SL without local weight sharing (P-SL)  \cite{pham2024splitlearninglocalweight}, which prevents adversaries from leveraging model inversion to reconstruct client data. Instead of sharing model parameters, this approach enables collaborative training without exposing private data. The gradient computation in this modified setting is given by:
\begin{equation*}
\nabla L(\text{outputs}, \text{labels}) =
\nabla_{u_i, w} L \left( g_w([z_i, z^{\text{cache}}]), [y_i^{\text{train}}, y^{\text{cache}}] \right)
\label{eq:GradientObfuscation}
\end{equation*}
where \( z_i = f_u(x_i) \) represents the smashed data from the current client, and \( z_{\text{cache}} \) consists of previously cached smashed data, effectively mitigating privacy leakage from individual updates.

An alternative strategy is to \emph{invert the SL setup}, wherein clients hold the labels while the server processes encrypted features, as proposed in \cite{CURE2024Kanpak}. This approach reduces the computational burden on resource-constrained clients while preserving model utility. 

Other examples include \cite{pham2024splitlearninglocalweight, Li2022ResSFL}, which explore communication and data-sharing regulations. Such protocol modifications ensure that client-side computations remain efficient while reducing the effectiveness of feature-space hijacking and reconstruction attacks \cite{turina2021fsl, Li2022ResSFL}. However, both architecture and protocol level modifications introduce a computational burden on the server, which must be managed to maintain scalability. 

While these methods can be empirically effective, they may incur a reduction in model accuracy and often lack formal security proofs, leaving them potentially vulnerable to minor adaptations in attack strategies. Key considerations include:
\begin{itemize}[nosep,leftmargin=*]
    \item \emph{Model Accuracy vs. Privacy}: Structural modifications may compromise performance.
    \item \emph{Robustness Under Diverse Attacks}: Without formal guarantees, even minor changes in attack tactics might bypass these defenses.
\end{itemize}

\subsubsection{Detection Mechanisms}
\label{sec:DetectionMechanisms}
Detection mechanisms focus on \emph{monitoring} the training process to identify malicious activity \emph{during or after} its occurrence, thereby enabling timely responses such as halting or rolling back updates. These methods are particularly relevant for scenarios involving overtly malicious adversaries who might engage in label-flipping, backdoor insertion, or model hijacking. Two primary detection approaches are employed:

\paragraph{\ref{sec:DetectionMechanisms}.1 Gradient/Update Monitoring:}
\label{sec:Gradient/Update Monitoring:}
This method involves analyzing the gradients or updates exchanged between clients and servers to detect abnormal variations. Outlier detection techniques can highlight significant deviations that may signal an ongoing attack.
Several studies have explored gradient anomaly detection to identify adversarial activity in SL \cite{Pinocchio, erdogan2022splitguard,erdogan2024splitout}. Considering the approach in~\cite{Pinocchio}, the detection score $DS_n$ is computed as follows:
\begin{equation}
DS_n = Sig(\lambda_{ds} (\hat{G}_n \cdot \hat{E}_n \cdot \hat{V}_n - \alpha))
\label{eq:GradientMonitoring}
\end{equation}
where \( \hat{G}_n \) represents the adjusted gradient similarity gap, \( \hat{E}_n \) is the logarithmic transformation of the polynomial approximation error, and \( \hat{V}_n \) quantifies the overlap between same-class and different-class gradients. \( Sig \) is a sigmoid activation that normalizes the detection score to the range \([0,1]\). This approach enables the identification of adversarial gradient manipulations by analyzing deviations in expected gradient behavior.

%investigate methods for monitoring gradient updates and detecting inconsistencies that could indicate training hijacking, backdoor attempts, or FSHA.

\paragraph{\ref{sec:DetectionMechanisms}.2 Weight Anomaly Detection:}
\label{sec:Weight Anomaly Detection:}
Some frameworks track model weight evolution, comparing it to normal behavior to detect anomalies and prevent adversarial tampering. Studies such as \cite{safeSplit2025rieger} analyze changes in weight distributions to detect anomalies, particularly those introduced by backdoor attacks, making them effective in identifying malicious training behaviors. One such metric for anomaly detection is the \emph{Rotational Distance} metric introduced in \cite{safeSplit2025rieger}, which quantifies the directional change in parameter space between consecutive training iterations:
\begin{equation}
\theta(t) = \arccos \left( \frac{\mathbf{B}_t \cdot \mathbf{B}_{t-1}}{\|\mathbf{B}_t\| \|\mathbf{B}_{t-1}\|} \right)
\label{eq:WeightAnomaly}
\end{equation}
where \( \mathbf{B}_t \) and \( \mathbf{B}_{t-1} \) are the backbone parameters at consecutive training steps, and \( \|\cdot\| \) denotes the Euclidean norm. The angular displacement \( \theta(t) \) measures the directional shift in model updates. However, backdoor attacks introduce abnormal directional changes in weight updates, causing \( \theta(t) \) to deviate from normal training patterns. By averaging pairwise differences in \( \theta(t) \) across clients, anomalies can be detected and flagged as potential threats.

%An important aspect of detection mechanisms is that they were initially introduced to counter FSHA-like reconstruction attacks. Over time, they have evolved alongside advancements in attack strategies. As both offensive and defensive techniques have progressed, detection mechanisms have expanded to identify and mitigate model inversion, backdoor, training hijacking, and poisoning attacks, as outlined in Section \ref{sec: Attack Tecniques Employed} and illustrated in Figure~\ref{fig:techniques considered against attacks}.  

% \begin{figure}
%     \centering
%     \includegraphics[width=\linewidth]{Defense/Attacks considered against techniques.png}
%     \caption{Defense techniques considered against attacks}
%     \label{fig:techniques considered against attacks}
%         \vspace{-1em}
% \end{figure}

Monitoring techniques have been refined to strengthen security while maintaining training integrity and robustness. Rieger et al. \cite{safeSplit2025rieger} propose periodically auditing trained models to validate participant honesty. They highlight the risk of falsely labeling honest participants as malicious, which could disrupt the training process. Their approach ensures that training remains robust while minimizing false positives in adversary detection. %Additionally, various other works~\cite{erdogan2023defense, erdogan2024splitout, Pinocchio} have explored gradient-based detection methods to distinguish between malicious and honest updates. These methods analyze anomalies in gradients sent by a potentially malicious server, leveraging outlier detection techniques to identify and filter adversarial updates.

% \begin{table*}[h!]
% \centering
% \renewcommand{\arraystretch}{0.5} % Adjust row height for readability

% \begin{tabular}{|c|c|c|c|c|l|}
% \hline
% \rowcolor[HTML]{EAEAEA} 
% \textbf{Defense Technique} & \textbf{Accuracy} & \textbf{Privacy} & \textbf{Computational Efficiency} & \textbf{Adversarial Model} & \textbf{References} \\ \hline
% Secure Computation & \ding{108} & \ding{108} & \ding{109} &  \ding{119} & \cite{khan2024make, SplitWithoutALeak,CURE2024Kanpak,khan2023more_secure_split,splitHE}\\  \hline
% Data Perturbation & \ding{119} & \ding{119} & \ding{108} & \ding{119} &  \cite{PrivateMail, pham2024enhancingaccuracyprivacytradeoffdifferentially, gawron2022feature, practical2021}\\ \hline
% Architectural Modification & \ding{119} & \ding{119} & \ding{108} & \ding{119} & \cite{pham2022binarizing, Abuadbba2020, pham2024splitlearninglocalweight, Li2022ResSFL, PSLF2023Wan, mao2023securesplit, NoPeek, DISCO, turina2021fsl, khowaja2024foesfooled} \\ \hline
% Gradient Detection & \ding{108} & \ding{119} & \ding{119} & \ding{108} & \cite{safeSplit2025rieger}\\ \hline
% Weight/Anomaly Detection & \ding{108} & \ding{119} & \ding{119} & \ding{108} & \cite{erdogan2024splitout, erdogan2023defense, Pinocchio}\\ \hline
% \end{tabular}

% \caption{Comparison of Defense Techniques in Split Learning. \ding{108}\space indicates strong presence, \ding{119} \space indicates partial presence, and \ding{109}\space indicates low effectiveness.}
% \label{tab:defense_techniques}
%     \vspace{-1em}
% \end{table*}

% We summarize aforementioned techniques in Table~\ref{tab:defense_techniques}.

\subsection{Defense Constraints}
\label{sec:DefenseConstraints}
Defense strategies in SL operate under constraints that affect how, and to what degree, they can be deployed: (i) assumptions about adversaries, % (Section~\ref{sec:ThreatModelConstraints}), 
(ii) requirements for verifying correctness when multiple parties are involved, % (Section~\ref{sec:CrossPartyVerificationConstraints}), 
(iii) the architectural details and training protocol configurations, % (Section~\ref{sec:ArchitecturalConstraints}),
and (iv) practical limitations in computational power or system availability. % (Section~\ref{sec:ComputationalOperationalConstraints}).

\paragraph{Threat Model Constraints}
\label{sec:ThreatModelConstraints}
Alongside the attacks discussed in Section~\ref{sec:DefenseAttackCoverage}, defense strategies differ in how they address adversaries that may only observe and infer (semi-honest) versus those that actively alter the environment (malicious). For semi-honest scenarios, cryptographic methods~\cite{CURE2024Kanpak,khan2024make,SplitWithoutALeak,SplitWays2023,splitHE}, data perturbation~\cite{PrivateMail,Pham2024Enhancing,practical2021}, and architecture-based approaches~\cite{pham2022binarizing,Abuadbba2020,mao2023securesplit,DISCO,Li2022ResSFL} generally suffice, as they aim to mitigate reconstruction attacks. In malicious settings, verification-based detection strategies~\cite{Pinocchio, erdogan2023defense,erdogan2024splitout,safeSplit2025rieger} are often necessary to detect and prevent adversarial behaviors such as poisoning or backdoor attacks. Balancing these approaches depends on the degree of trust among participants and the trade-off between privacy and performance overhead.

\paragraph{Multi-Party Verification Constraints}
\label{sec:CrossPartyVerificationConstraints}
When multiple clients or servers are involved, defenses may need enhanced validation mechanisms to differentiate between benign and malicious updates, as demonstrated in~\cite{erdogan2023defense,erdogan2024splitout,Pinocchio}. Some methods~\cite{safeSplit2025rieger,khowaja2024foesfooled} rely on cross-comparisons among clients, a trusted third party, or a server, which increases communication rounds and latency. In large-scale or bandwidth-limited environments, frequent verification may be impractical. Leveraging powerful nodes or third parties can mitigate some of these issues but introduces additional trust assumptions.

% In large-scale or bandwidth-limited environments, frequent verification may be impractical. The availability of powerful nodes or third parties can mitigate some of these issues, but it introduces its own trust assumptions.

\paragraph{Architectural Constraints}
\label{sec:ArchitecturalConstraints}
Certain defenses~\cite{pham2022binarizing,Abuadbba2020} depend on the model’s partitioning and the number of layers retained on the client side~\cite{pham2022binarizing} or specialized activation functions~\cite{mao2023securesplit}, can limit information leakage but may also reduce flexibility or performance on complex tasks. 
Architectural considerations become increasingly crucial when striving for a generalized approach that minimizes reliance on specific empirical setups. 

% There are also training methods that may not align well with any of these defense techniques or that could yield suboptimal results if model efficiency is sacrificed for privacy gains.

\paragraph{Computational and Operational Constraints}
\label{sec:ComputationalOperationalConstraints}
Practical deployments of SL, especially on edge or mobile devices, limit how often encryption, extra communication rounds, or gradient checks can be performed. Heavy cryptographic schemes may be infeasible, and much of the work in this domain focuses on improving computational efficiency~\cite{CURE2024Kanpak, SplitWithoutALeak,SplitWays2023}. Additionally, defenses that assume continuous client availability are prone to disruption if participants drop out or arrive late, as considered in~\cite{safeSplit2025rieger}. Some training modifications also limit the flow of information during learning~\cite{pham2024splitlearninglocalweight}. These operational constraints highlight the need for strategies that remain robust under unpredictable network conditions and resource limitations.

\subsection{Defense Effectiveness}
\label{sec:DefenseEffectiveness}

Defense effectiveness in SL is evaluated based on several key factors, including applicability to different attack types, assumed client-server setups, computational constraints, and tested model architectures. The alignment between defense strategies and attack vectors is crucial, as protection-based approaches such as cryptographic techniques, DP, and architectural modifications focus on preventing information leakage, while detection-based methods such as gradient monitoring and anomaly detection aim to identify and halt adversarial behavior.

\subsubsection{Attack Coverage and Defense Alignment}
\label{sec:DefenseAttackCoverage}
Each defense mechanism addresses specific adversarial threats in SL. \emph{Data perturbation defenses} \cite{Abuadbba2020, PrivateMail, Pham2024Enhancing,practical2021}, including DP and feature decorrelation, are effective against \emph{label inference and reconstruction attacks}, making it more difficult for adversaries to extract information from gradients or intermediate representations. However, these approaches often introduce a trade-off in accuracy. \emph{Secure computation techniques} \cite{CURE2024Kanpak, SplitWithoutALeak, LoveHate, splitHE, khan2024make}, such as HE and FSS, offer robust privacy guarantees by allowing encrypted computations.
%, but they introduce high computational overhead, making them impractical for real-time or resource-constrained scenarios.

\emph{Architectural modifications}\cite{pham2022binarizing, Abuadbba2020, pham2024splitlearninglocalweight, Li2022ResSFL, PSLF2023Wan, mao2023securesplit, NoPeek, DISCO, turina2021fsl, khowaja2024foesfooled}, including cut-layer adjustments and novel activation functions, significantly mitigate \emph{feature-space hijacking and reconstruction attacks}, but their effectiveness depends on the model’s depth and complexity.

Detection mechanisms are crucial for identifying adversarial actions during training. \emph{Gradient anomaly detection}~\cite{erdogan2024splitout,erdogan2023defense,Pinocchio} is effective against \emph{poisoning and hijacking attacks} by analyzing inconsistencies in the updates exchanged between clients and servers. \emph{Model weight anomaly detection} \cite{safeSplit2025rieger} is particularly useful against \emph{backdoor attacks}, as it can identify hidden manipulations within the model weights by monitoring deviations from expected parameters. The effectiveness of these mechanisms relies on accurately profiling normal training behavior, and they may be vulnerable to adversaries who subtly manipulate updates to evade detection.

\subsubsection{Client-Server Setup and Deployment Scenarios}
\label{sec:DefenseClientServerSetup}
The client-server setup influences the feasibility and effectiveness of defenses. In \emph{single-client SL (SCSL)}, cryptographic defenses such as HE and FSS are more feasible, as encryption and decryption overhead is limited to a single participant. Architectural modifications are also easier to apply, as there is no need for cross-client consistency. 

In \emph{multi-client SL (MCSL)}, on the other hand, \emph{client-side gradient monitoring and anomaly detection} become more relevant, as adversarial participants may be disguised among honest clients. \emph{Data perturbation techniques}, including DP, require careful tuning in MCSL, as variations in data distributions may affect privacy-utility trade-offs. In addition, DP alone are insufficient in terms of defense as discussed earlier (Section~\ref{sec: Attack Effectiveness}). \emph{Secure computation methods} face scalability challenges, as the overhead increases with an increasing number of clients. In MCSL, \emph{architectural modifications} such as split-layer depth optimization and secure aggregation become more practical, as computational burdens are distributed across multiple parties. Lastly, communication latency and encryption overhead remain key concerns for cryptographic techniques.

\subsubsection{Model and Dataset Considerations}
\label{sec:DefenseModelDataset}
Defense effectiveness also depends on the complexity of the model architecture and the dataset's properties. In \emph{lightweight models} such as convolutional neural networks for image classification, \emph{DP and gradient perturbation} are effective \cite{Pham2024Enhancing,practical2021}, as small distortions in features do not drastically impact performance. Cryptographic methods such as HE remain feasible in small-scale models but introduce significant latency \cite{CURE2024Kanpak, khan2023more_secure_split}. In \emph{deep architectures} such as ResNet and transformers, \emph{secure computation techniques} struggle due to their high computational complexity, while \emph{architectural modifications} \cite{Abuadbba2020,pham2022binarizing, mao2023securesplit} become more critical to mitigate feature-space hijacking. \emph{Detection-based methods}, such as gradient anomaly analysis, are highly dependent on the model architecture and training setup, as the information carried in gradients is significantly influenced by these factors. For instance, as discussed in~\cite{erdogan2024splitout}, larger gradients may suffer from the curse of dimensionality, since these methods rely on distance and correlation analysis among data.

%% file: takeaways.tex
\section{Key Observations and Takeaways}\label{sec:takeaways}
%In this paper, we have thoroughly examined SL attacks and defenses. 
We outline key observations derived from our analysis on SL attack and defenses below: 
\begin{itemize}[nosep,leftmargin=*]
    \item \textbf{Architecture:} The feasibility of attacks and defenses is heavily influenced by the architectural design (e.g., VanSL, USL) and data partitioning strategies (horizontal vs. vertical). We observe that most attack research focuses on VanSL, which is highly vulnerable to feature-space hijacking and model inversion due to the exposure of smashed data at a single cut layer. In contrast, USL and SFL face increased label exposure risks due to multi-hop data transmissions. Similarly, the majority of defenses focus on the VanSL setting (Figure~\ref{fig:Split learning type pie}). Furthermore, \revised{we observe that No-Label SL (NLSL) and Multi-hop SL (MHSL) are the least explored variants in the literature despite their distinct privacy and trust characteristics. %, which offer new defense opportunities and unique vulnerabilities. 
    NLSL enhances label privacy by keeping labels on the client, requiring thorough analysis to assess its resistance to inference attacks targeting label information via gradients or activations. MHSL, with its chained computation across multiple entities, has received minimal attention in the literature, and its unique multi-party computation and trust assumptions remain inadequately characterized. These observations directly address \textit{RQ1}, which seeks to understand how adversarial objectives can be systematically taxonomized in relation to architectural vulnerabilities in SL. Our findings underscore the need to extend taxonomies beyond VanSL to capture under-explored variants with distinct structural vulnerabilities.}
    \revised{\item \textbf{Semi-Honest Adversaries:} Our taxonomy reveals a critical reliance on the semi-honest adversarial model, where adversaries passively follow the protocol. While this simplifies threat modeling, it underestimates real-world risks, as attacks like label flipping, gradient manipulation, or collusion exceed the assumptions of this model, rendering many defenses ineffective against stronger adversaries per \textit{RQ3}. %As illustrated in Figure~\ref{fig:Adversarial Model} and discussed in Section~\ref{sec:Defense Mechanisms}, the prevalence of this assumption creates a research gap of defenses that are empirically effective against passive leakage may break under active manipulation. 
    }
    \revised{\item \textbf{Generalization of Attack Mechanisms:} Several core attack strategies, especially those that exploit the cut layer bottleneck, such as latent representation reconstruction~\cite{Unleashing_tiger, mao2023securesplit} and property inference based on intermediate activations~\cite{Unleashing_tiger}, demonstrated effectiveness across a variety of SL models and datasets. This suggests that the act of splitting and transmitting intermediate representations introduces consistent privacy vulnerabilities that are not easily mitigated by simply changing model architecture or task. These findings respond to \textit{RQ2}, which seeks to understand the generalizability and stealth of SL attack mechanisms. The persistence of these attacks across different conditions suggests an architectural invariance in SL's threat surface. This highlights both a concern and an opportunity: to develop model-agnostic defenses that reduce information leakage in representations $z$ and gradients $\nabla$ as discussed in Section~\ref{Label Inference Attacks}, for example, via dimensionality reduction. }
    \revised{\item \textbf{Inherent Risk of the Cut Layer:} The interface between the client-side and server-side models, where $z$ or $\nabla$ are exchanged, consistently emerges as a critical attack surface in SL. Multiple attacks, including FSHA~\cite{Unleashing_tiger}, model inversion via gradients~\cite{erdougan2022unsplit}, GAN-based input reconstruction~\cite{zeng2025gan}, and various label or property inference strategies~\cite{kariyappa2023exploit, Unleashing_tiger,erdougan2022unsplit}, exploit this specific interface to extract sensitive information. This highlights the need to prioritize specifically the security of the cut layer. Moreover, it underscores that defenses focused solely on obfuscating raw inputs, such as local DP (see Section~\ref{sec: Attack Effectiveness}), are insufficient, as critical information is often preserved and leaked through $z$ and $\nabla$. Effective SL defenses must directly mitigate the information content traversing this unavoidable interface between model partitions, which directly relates to \textit{RQ2’s} examination of attack vectors and vulnerable interfaces.}
    \revised{\item \textbf{Reliance on Strong Adversarial Capabilities:} Many prominent SL attack studies assume strong adversarial capabilities, such as full or partial model access~\cite{Unleashing_tiger,zeng2025gan} or access to auxiliary datasets aligned with the training distribution~\cite{zhang2024functionality,ismail2023analyzing,yu2023backdoor}. While useful for exposing theoretical vulnerabilities, these assumptions often diverge from practical deployment scenarios, where adversaries may have limited knowledge, access, or resources. 
    In addressing \textit{RQ3}, we systematically differentiate between strong and constrained adversarial assumptions across attack studies, enabling more accurate threat modeling and highlighting the need for context-aware, practically-deployable defenses.}
    \revised{\item \textbf{Leakage Persistence Beyond Gradients:} Studies have demonstrated that intermediate activations exchanged during the forward pass can leak sensitive information, enabling label inference~\cite{liu2024similarity,zhu2023passive} and data reconstruction~\cite{mao2023securesplit, Unleashing_tiger} even in the absence of gradient information. This critical finding shows that SL remains vulnerable even during inference-only deployments or when gradient leakage is mitigated through secure aggregation or other protective mechanisms. This emphasizes the need for defenses targeting information leakage in $z$ —such as its semantic consistency or correlation with inputs— beyond just the backward pass, addressing \textit{RQ1} on the nature and sources of information leakage in SL.}
%     \item \textbf{Protection-Based Defenses.} Data perturbation techniques like DP modify data or intermediate representations to obscure sensitive information.
% Secure computation techniques rely on cryptographic methods enabling computations on encrypted data while preserving privacy. 
% Architecture Modification methods such as shrinking the client-side cut layer and specialized activation functions structurally limit exposed information during training.
%     \item \textbf{Monitoring-Based Defenses. } Gradient \& update monitoring tracks gradient evolution to detect backdoor insertion, model inversion, and poisoning attacks while weight anomaly detection monitors model weight changes over time to identify deviations from expected behavior.
%
%     \item \textbf{Privacy vs. Efficiency of Defense Strategies:} Secure computation (e.g., HE) provides strong privacy guarantees but imposes high computational and communication costs. Thus, resource-constrained clients struggle to implement cryptographic and obfuscation techniques without sacrificing model performance. Monitoring and architecture modifications, on the other hand, offer practical solutions but lack rigorous security proofs compared to cryptographic methods.
%
    % \item \textbf{Deployment:} SL remains vulnerable to various privacy attacks, necessitating comprehensive defense mechanisms. Existing defenses provide partial solutions, but integrating scalable, low-overhead techniques is crucial for real-world SL deployment.
    %
\revised{
    \item \textbf{Privacy--Utility--Overhead Trade-off:}
SL defenses show a clear three-way trade-off between privacy, model utility (e.g.\ accuracy), and computational or communication overhead. Methods range from lightweight approaches with weaker guarantees to heavy cryptographic schemes with high costs; while DP offers a tunable middle ground whose optimal balance is still application-specific and challenging to achieve. Homomorphic Encryption (HE) and Function Secret Sharing (FSS) offer strong theoretical privacy by enabling computation on encrypted data. However, their high computational and bandwidth overheads restrict practical use to simple models or scenarios, where privacy outweighs performance.
This addresses \textit{RQ4} by discussing the place of cryptographic techniques in the landscape of defense techniques. The three-way trade-off is addressing \textit{RQ5}, which discusses the suitability, constraints, and potential optimization of defenses for specific tasks, and also connects to \textit{RQ6} regarding what constitutes a \emph{successful} defense in practice, often involving a balance of these factors.}
%
%\revised{\item \textbf{Crypto vs Reality:}
%Homomorphic Encryption (HE) and Function Secret Sharing (FSS) offer strong theoretical privacy by enabling computation on encrypted data. However, their high computational and bandwidth overheads restrict practical use to simple models or scenarios, %such as general PPML, 
%where privacy outweighs performance.
%This addresses \textit{RQ4} by discussing the place of cryptographic techniques in the landscape of defense techniques. Also, observing that cryptographic techniques offer strong theoretical guarantees but are constrained by high computational and bandwidth costs (making them impractical for complex models or resource-limited settings), we address \textit{RQ5} by indicating that these techniques are suitable for scenarios where machines possess high computational resources and the optimization concerns in this techniques.}
%
% \revised{\item \textbf{Differential Privacy:}
% DP provides a formal, quantifiable privacy guarantee ($\epsilon$-DP) with lower computational costs than full cryptography. It's calibrated noise, however, directly trades accuracy for privacy: lowering $\epsilon$ means more noise and, as many studies show, a noticeable drop in utility. Tuning the right noise for each SL task and data type remains a key open issue.}
%
\revised{\item \textbf{Architectural Changes:}
Modifying SL —by shifting the cut layer or shrinking the shared 'smashed data'— can reduce exposure and increase attack difficulty, but studies agree that these changes alone rarely ensure full or provable privacy, serving best as complements to stronger defenses. This observation aligns with \textit{RQ4's} investigation into the strengths and limitations of current defense strategies in mitigating information leakage. Architectural changes are a good fit for situations where one does not have much computing power or when one wants to add a basic, extra layer of security alongside stronger methods.}
%As \textit{RQ4} also considers the limitations of defenses, a key downside is that they rarely offer full or guaranteed privacy on their own, this answers \textit{RQ5}. 
%
%
%\revised{Our analysis for \textit{RQ4} identifies key categories of defense techniques aligned with the taxonomy: cryptographic methods (e.g., HE, FSS), perturbation-based approaches (e.g., DP), architectural modifications, and detection strategies. As discussed in Section~\ref{sec:Defense Mechanisms} and our takeaways, these are evaluated in response to \textit{RQ5} regarding their suitability, limitations, and real-world applicability.}
%
%
\revised{\item \textbf{Confidentiality \& Integrity:}
Most SL defenses target two goals: confidentiality (protecting data, labels, or attributes) and integrity (preventing model manipulation with monitoring or robust aggregation). Availability remains under-explored in the reviewed literature. This distinction between defense goals directly informs \textit{RQ6} —how techniques prioritize confidentiality vs. integrity— and supports the taxonomy outlined in \textit{RQ4}; notably, detection strategies, crucial for identifying active threats like poisoning or backdoors (though reliant on robust baselines and unable to address passive leakage), fall under this discussed class and thus also address aspects of \textit{RQ4}.}
\revised{\item \textbf{Hybrid Defenses:}
While layering defense techniques may seem promising, the literature highlights complex interactions—such as applying DP to encrypted data or combining noise with architectural constraints, which can increase overhead, obscure true privacy guarantees, or introduce new vulnerabilities. These findings, central to \textit{RQ5}, emphasize that effective composition requires careful joint analysis, as naive stacking is often inadequate.}
\revised{\item \textbf{Prominence of Studied Defenses:} The literature shows DP as the most extensively studied approach for enhancing confidentiality in SL, followed by HE. In contrast, detection research focuses primarily on gradient analysis. Highlighting these commonly used techniques —DP, HE, and gradient analysis— directly informs \textit{RQ4} on prevalent SL defense strategies.}
\end{itemize}

\section{Open Research Directions}\label{sec:Open Research Directions}
Building on our key observations, we outline the key open research directions below:
\begin{itemize}[nosep,leftmargin=*]
    % \item \textbf{Advanced Attack and Defense Strategies.} Future work can explore novel attack and defense strategies extending beyond the VanSL setting to target more complex SL variants. 
    %
    \item \revised{\textbf{Advanced Attack and Defense Strategies:} A key research direction emerging in response to \textit{RQ1} involves extending adversarial taxonomies and corresponding defense mechanisms to SL variants beyond the conventional VanSL setting. Alternative SL variants such as USL, Hybrid-SL, NLSL, and MHSL exhibit distinct architectural patterns (e.g., multi-hop communication, hybrid cut points) and data handling procedures, which introduce previously unaddressed attack surfaces. As outlined in Section~\ref{sec: Threats and Attack Scenarios} and Section~\ref{sec:Defense Mechanisms}, these non-standard configurations may necessitate variant-specific defense strategies that are not adequately covered by existing VanSL-centric approaches.}
    \revised{Alongside exploring novel approaches, techniques from the broader PPML literature could be adapted to SL when there are matching problem structures. For instance, adaptive differential privacy used in federated learning~\cite{fu2022adap,talaei2024adaptivedifferentialprivacyfederated,Chen2024} could be incorporated to dynamically adjust noise levels based on model updates. This adjustment could help balance the trade-off between model accuracy and privacy by reducing noise in scenarios where updates pose a lower privacy risk.}
    \revised{\item \textbf{Broadening Threat Models Beyond Semi-Honest Adversaries:} 
    Future research should focus on developing a progressive adversarial framework that evaluates defenses across a spectrum of adversarial strengths, from semi-honest actors to fully malicious colluding parties. SL systems are especially vulnerable to adversaries capable of poisoning gradients, manipulating activations, or tampering with training states. We advocate for hybrid evaluation protocols that integrate empirical testing with theoretical analysis to assess robustness. Furthermore, adopting formal models from cryptographic literature (e.g., Byzantine threat models~\cite{fang2020local}, covert adversaries~\cite{aumann2010security}, or adaptive adversaries~\cite{arora2012online}) can help bridge the current gap between theoretical security guarantees and practical adversarial capabilities, thus advancing the scope of \textit{RQ3} on evaluating the realism and comprehensiveness of adversarial assumptions.}
    \revised{\item \textbf{Toward Uncertainty-Aware Defense Design:} A promising research direction is the development of uncertainty-aware defenses via entropy analysis of intermediate representations in SL systems. Low entropy in smashed data —often caused by deterministic or sparse activations— correlates with increased vulnerability to inference attacks, as it signals reduced uncertainty exploitable by adversaries. Operations like ReLU and max-pooling compress representational diversity, aligning features closely with input semantics and elevating privacy risk. In contrast, randomized activations and dropouts introduce beneficial uncertainty that hinders inference. Formalizing these effects using information-theoretic tools like mutual information and representation entropy could quantify the trade-off between expressivity and leakage. Additionally, comparing differential entropy before and after the cut layer may help identify optimal cut points that balance utility and privacy. Entropy-based metrics thus offer a principled foundation for designing more privacy-resilient SL architectures, contributing to \textit{RQ2} on structural factors that shape the effectiveness of inference attacks in SL.}
    %
    % \revised{\item \textbf{Toward a Deeper Mechanistic Understanding and Protocol-Specific Attacks:} To address RQ2, future research should move beyond empirical attack demonstrations and explore why certain attack classes—such as GAN-based reconstructions, gradient inversion, and data poisoning—are particularly effective in SL. This requires systematic studies that isolate the impact of architectural factors like cut layer depth, activation functions, and gradient flow on attack success. A key direction is to examine whether these attack generalizations can be theoretically linked to structural aspects of SL, such as the information bottleneck introduced by the split, enabling a more principled attack taxonomy. Additionally, there is scope for uncovering new attack vectors that exploit protocol-specific SL vulnerabilities, including timing side channels, adaptive inference queries, server-side aggregation tampering, or deterministic communication patterns. Such work would help expand and sharpen the taxonomy of SL-specific threats.}
    \revised{\item \textbf{Evaluating Practical Threats Under Realistic Conditions:} In response to \textit{RQ3}, future research should prioritize the development and evaluation of attacks under more constrained and realistic conditions. This includes black-box settings, limited or mismatched auxiliary data, bounded query or compute budgets, and scenarios involving partial or probabilistic knowledge of the target model. Crucially, quantifying how attack effectiveness degrades under practical constraints is vital to building accurate and actionable threat models for real-world SL deployments.}
%
    % \item \textbf{Enhancing Defenses in Malicious Multi-Client Setups: } Future research should address coordinated adversarial behavior while improving computational efficiency. Exploring zero-knowledge proofs (ZKP) can be a potential cryptographic alternative for SL security.
    % \item \textbf{Advancing Holistic Monitoring Strategies:} Developing proactive defense mechanisms to counter a broader range of SL attacks is an interesting research direction. Integrating insights from federated learning, secure inference, and side-channel attacks (discussed in \cite{debenedetti_sidechannel}) can improve robustness in SL.
    \revised{
    \item \textbf{Advancing Holistic Monitoring Strategies:} Future research should prioritize proactive, multi-layered defenses to address SL's broad attack surface: a direction already emerging in recent work, as noted in Section~\ref{sec:DefenseTechniquesEmployed}. Integrating anomaly detection (e.g., gradient or behavior-based) \cite{erdogan2024splitout} with insights from federated learning —such as techniques developed to achieve secure aggregation for verifying client updates \cite{karakocc2024fault}— can enhance detection capabilities without significant overhead. Additionally, defenses against side-channel threats (e.g.,~\cite{debenedetti_sidechannel}) and secure inference techniques can strengthen robustness against covert attacks. Hybrid approaches may be necessary —combining secure computation methods (for confidentiality of client data) with monitoring (for learning integrity)— to realize a truly holistic defense strategy in SL. This means that monitoring mechanisms should be integrable with other defense techniques.}
    \revised{\item \textbf{Instance Encoding and Potential Threats:} Intermediate representations in SL —often referred to as smashed data— can be viewed as encodings that have the potential to leak sensitive information, even without full input reconstruction. This issue is known as the \textit{instance encoding problem}~\cite{carlini2021encoding,maeng2023fisher}. Carlini et al.\cite{carlini2021encoding} reframed the PPML problem as instance encoding, analyzing the statistical underpinnings of privacy in ML, comparing techniques, and offering bounds and empirical results under specific attacks. Maeng et al.\cite{maeng2023fisher} introduced a Fisher information based framework to quantify and bound leakage in PPML, also leveraging the instance encoding perspective. Given the alignment in goals and definitions, this concept provides valuable insights for improving the security of SL systems.}
    \revised{\item \textbf{Verifiable Training and Inference:}
    Ensuring the verifiability of training and inference in SL, alongside privacy, is another crucial objective. Zero knowledge (ZK)-proofs enable parties to prove correct computation without leaking unauthorized information, offering a promising means to verify the correctness of training in PPML. This capability is vital where malicious participants alter training data, introduce backdoors, or poison the training process. Peng et al.~\cite{peng2025survey} survey the use of ZK-proofs in machine learning, categorizing approaches into verifiable training, testing, and inference, and proposing an abstract framework for their application in ML contexts, which can be beneficial for SL settings. Efficiency would constitute a crucial concern when ZK-proofs are employed.}
    \item \textbf{Exploring Hardware-Based Security Solutions:} Hardware-based solutions, such as trusted execution environments (e.g.,~\cite{huang2024TEE}), are under-explored and show promise for enhancing security and privacy in SL.
    \revised{\item \textbf{Resilience to Distributed System Faults} Ensuring the training process remains robust and continues effectively despite common distributed system faults, such as client dropouts, late-arrivals, and potential communication errors. Strategies focus on maintaining training integrity and progress even when parts of the system fail.}
\end{itemize}

%% file: related.tex
\section{Related Work}\label{sec:related}
Several Systematization of Knowledge (SoK) papers have examined privacy-preserving methodologies across collaborative learning paradigms. Podschwadt et al.~\cite{podschwadt2021sok} presented a systematization of deep learning to preserve privacy utilizing HE, focusing on the computational complexities and practical limitations of HE-based solutions. Ng and Chow~\cite{ng2023sok} conducted an extensive SoK examining cryptographic approaches to privacy preservation in deep learning. Mansouri et al.~\cite{mansouri2023sok} systematically analyzed secure aggregation techniques for FL, categorizing encryption-based and MPC-based approaches while evaluating their performance, scalability, and resilience against adversarial attacks.

While FL and SL share distributed learning characteristics, SL presents distinct challenges arising from its architectural design, particularly in the transmission of intermediate activations between participants. Conventional secure aggregation methods developed for FL cannot be directly applied to SL without substantial modifications, as SL's vulnerabilities emerge from unique adversarial capabilities, including feature-space leakage and inference attacks.
In examining SL-specific security concerns, Pham and Chilamkurti~\cite{pham2023data} surveyed data leakage threats, identifying gradient leakage, label inference, and feature reconstruction attacks. Hu et al.~\cite{hu2025review} conducted a review and experimental evaluation of SL, highlighting the variability in SL paradigms concerning cut-layer selection, model aggregation, and label sharing. While these works provide highly valuable empirical insights, we uniquely and systematically categorize security and privacy challenges in SL, establishing a novel formal taxonomy of attack vectors and mitigation techniques, analyzing their effectiveness and limitations. 

%To the best of our knowledge, no prior work has systematically categorized SL's security and privacy challenges. Existing studies explore attacks and defenses but lack a unified framework. We bridge this gap by structuring attack strategies and mitigation techniques to clarify their effectiveness and limitations. 
% Our work provides the first systematic classification of SL threats and defenses, organizing them by employed techniques, constraints, and effectiveness.

% Although cryptographic solutions such as Secure Multiparty Computation (MPC) and Homomorphic Encryption (HE) have been explored for privacy-preserving SL, existing research lacks a structured comparative analysis of their effectiveness against various adversarial models. Our work addresses these limitations by providing the first systematic classification of attack strategies and defense mechanisms in SL, organizing threats based on their methodologies, effectiveness, and operational constraints.

%% file: conclusion.tex
\section{Conclusion}
We systematically explored the security and privacy landscape of Split Learning (SL) by categorizing various attack strategies and defense mechanisms. We presented a novel taxonomy of attacks and defenses in SL,
categorizing them along: (i) strategies, (ii) constraints, and (iii) effectiveness. While SL presents a promising framework for distributed/outsourced machine learning, its inherent vulnerabilities -—ranging from data reconstruction and label inference to adversarial manipulation—- underscore the need for robust countermeasures. Existing defenses --including cryptographic techniques, differential privacy, and architectural modifications-- often introduce trade-offs in computational efficiency and scalability. Employing the key observations and takeaways presented in this paper, %By advancing our understanding of attack surfaces and defense strategies, 
future research can enhance the resilience of SL, making it a viable solution for secure collaborative learning.
\section*{Acknowledgements}
We acknowledge TÜBİTAK (the Scientific and Technological Research Council of Türkiye) project 124N941. 
The authors utilized ChatGPT-4o~\cite{openai2024gpt4o} to refine the text in Sections 1, 4, and 5 for improving text, shortening sentences, correcting typos and grammatical errors to enhance readability and clarity.

%% file: supplementary.tex
\section{Additional Split Learning Variants}\label{slvariants}
Here, we explain the remaining SL variants, namely multi-hop and no-label split learning. 
\par\noindent\textbf{Multi-hop Split Learning (MHSL):}\label{MHSL}
MHSL extends the SL concept to multiple parties, creating a chain of computation across multiple entities. The model is divided into several segments $(f_{c_1}, f_{s_1}, f_{s_2},...,f_{c_n})$ where the subscript denotes the entity processing that segment. The client with input $x$ begins by computing $(z_{c_1} = f_{c_1}(x))$ and sends $z_{c_1}$ to the first server. Each subsequent server $i$ computes $z_{s_i} = f_{s_i}(z_{s_{i-1}})$ and forwards it to the next entity. The final client computes the output $\hat{y} = f_{c_n}(z_{s_{n-1}})$ and calculates the loss. During backpropagation gradients flow in the reverse direction through the chain. %Each entity computes the gradients for its parameters and the gradients with respect to its input, passing the latter to the previous entity in the chain. 
While the ordering of clients and servers in MHSL is flexible, there are practical constraints. A client must process the first model segment to handle the $x$, and the last entity whether a client or a server must compute the loss. Intermediate entities can be either servers or clients, depending on computational needs. For example, in a medical collaboration setting, hospitals can process the initial and final model segments while cloud servers handle intermediate computations. In contrast, in a business forecasting scenario, multiple companies may preprocess data before sending it to a server for final inference. This flexibility allows MHSL to adapt to various distributed learning scenarios while ensuring privacy and scalability.
\par\noindent\textbf{No-Label Split Learning (NLSL):}\label{NLSL}
NLSL is designed to keep both $x$ and $y$ private to the client. In NLSL, the client processes $f_c$ to compute $z_c$ and sends it to the server. The server computes the forward pass through the remaining layers $f_s$ to generate $\hat{y}$. However, rather than calculating the loss on the server, the server sends $\hat{y}$ back to the client. The client then computes the loss $L(y, \hat{y})$ using the true labels $y$ and calculates the gradient $(\nabla_{\hat{y}} = \frac{\partial L}{\partial \hat{y}})$. The client sends $\nabla_{\hat{y}}$ back to the server, which begins the backpropagation process by computing $(\nabla_{z_c} = \frac{\partial L}{\partial z_c})$ and sending it to the client. Finally, the client updates the parameters of $f_c$ using the received gradients. Similar to USL, NLSL ensures that the client is responsible for computing the loss, preventing the server from having direct access to $y$. However, unlike USL, where the client also processes the final layers after receiving intermediate representations from the server, in NLSL, the server completes the entire forward pass, ensuring that only the client handles label-related computations, offering stronger privacy guarantees.
\section{Details of Data Reconstruction Attacks} \label{sec:Details of Data Reconstruction Attacks}
Data reconstruction attacks in split learning exploit smashed data \( z_c \) to infer private client inputs. They are primarily categorized into (i) Feature Space Hijacking Attack (FSHA), (ii) Model Inversion Attack, (iii) Functionality Stealing, (iv) Generative Adversarial Network (GAN), and (v) Feature Reconstruction techniques. Each technique exploits distinct aspects of the model behavior to compromise training and data privacy as explained below: 
\paragraph{i. Feature Space Hijacking Attack (FSHA):}
A prominent data reconstruction attack is FSHA~\cite{Unleashing_tiger,gawron2022feature}, which allows a malicious server to reconstruct the private data in a VanSL setup. FSHA consists of three key components: (i) Pilot network ($\tilde{f_c}$): dynamically defines the target feature space $\tilde{z_c}$ and is responsible for mapping between raw input $x$ and $\tilde{z_c} = \tilde{f_c}(x)$ for the client network. (ii) Inverse network ($\tilde{f_c}^{-1}$): trained to approximate the inverse function of $\tilde{f_c}$, allowing the malicious server to reconstruct $x$ from $\tilde{z_c}$, and (iii) Discriminator ($D$): an adversarially-trained network that indirectly guides to learn the mapping between $x$ and $\tilde{z_c}$. FSHA follows a two-step adversarial training process. In the first step, the server begins by sampling a batch from public dataset $x_{pub}$ to train $\tilde{f}_c$ and $\tilde{f_c}^{-1}$, ensuring the networks converge by minimizing the reconstruction loss:
\begin{equation}
    L_{\tilde{f_c}, \tilde{f_c}^{-1}} = d(\tilde{f_c}^{-1}(\tilde{f_c}(x_{pub}), x_{pub})
\end{equation}
where $d$ is a suitable distance function, such as Mean Squared Error (MSE). In the second step, the server adversarially trains the $D$ loss to distinguish between $\tilde{z_c}$ from the one induced from the $z_c$:
\begin{equation}
    L_D = \log(1 - D(\tilde{z_c})) + \log(D({z_c}))
\end{equation}
After each local training step for $D$, the malicious server can then train the network by forging the gradient using $D$ to reconstruct an adversarial loss function for $f_c$: $L_{{f_c}} = \log(1 - D({z_c}))
$.
After adversarial training, the server can reconstruct the client’s private input using: $
    \tilde{x} = \tilde{f_c}^{-1}({f_c}(x))$
where $\tilde{x}$ is the reconstructed approximation of the client’s original data. 
\paragraph{ii. Model Inversion:}
Another key attack technique in data reconstruction is Model Inversion~\cite{he2019model}, applied in~\cite{erdougan2022unsplit}, which exploits the relationship between the gradients and the original input in a VanSL setting. In this approach, the honest-but-curious server iteratively optimizes both the original input $x$ and the client model parameters $\theta$ to reconstruct the data. Instead of explicitly matching the gradients, the attacker performs coordinate gradient descent, where the optimization alternates between updating $x$ and $\theta$ to minimize the mean squared error (MSE) for both input and parameter updates. The attack minimizes the following objective:
\begin{equation}
    \tilde{x} = \arg\min_{\tilde{x}^*} MSE (\tilde{f_c^*}({\tilde{x}^*,\tilde{\theta}^*}),{f_c}({x,\theta})) + \lambda \operatorname{TV}(\tilde{x}^*)
\end{equation}
\begin{equation}
   \tilde{\theta} = \arg\min_{{\tilde{\theta}^*}} MSE (\tilde{f_c^*}({\tilde{x}^*,\tilde{\theta}^*}),{f_c}({x,\theta})) 
\end{equation}
Here, $\tilde{f_c^*}$ represents the random initialization of the client network, \( \operatorname{TV}(\tilde{x}^*) \) is the Total Variation~\cite{rudin1992nonlinear} to enforce smoothness, and \( \lambda \) is a regularization parameter. The attack is demonstrated in a single client single server (SCSL) setup and it can extend to multi-client configurations through two key properties: sequential client training and shared parameter updates. Since the server interacts with only one client at a time and all clients update the same parameter set, a multi-client setup functionally mirrors a single-client configuration with aggregated data.
\paragraph{iii. Functionality Stealing:}
The functionality stealing attack in~\cite{zhang2024functionality} allows a semi-honest server to train a \textit{pseudo-client model} (\( \tilde{f}_c \)) that closely mimics $f_c$ without knowing its structure in VanSL and USL setups. The server achieves this by using intermediate server models and a small set \( x_{{pub}}\) of size $N$. %, where $\quad n \in [0, N]$. 
The goal is to make $\tilde{f}_c$ learn the mapping, making it functionally identical to \( f_c \), while training the server model separately. Mathematically, the training objective for the pseudo-client model is to minimize the KL-Div (Kullback-Leibler  Divergence), which measures the difference between the soft labels of \( \tilde{z_c} \) and \( z_c \) rather than the hard labels.
% \begin{equation}
%     L_{\text{KL-Div}} = \text{KLDivergenceLoss}(\text{soft}(\tilde{z_c}^n), \text{soft}({z_c}^N))
% \end{equation}
Once the functionality is stolen, the server trains a reverse mapping function \( f_s^{-1} \) to transform the feature space of smashed data back into the original input space. Since the pseudo-client model has already learned to produce feature representations similar to \( f_c \), the server can fine-tune \( f_s^{-1} \) to improve reconstruction accuracy. The server can apply this attack to multi-client SL directly without any modification.
\paragraph{iv. Generative Adversarial Network (GAN):}

A data reconstruction attack using GAN~\cite{mao2023securesplit} is performed in a VanSL setup, where a semi-honest server trains a generator \( G \)  to synthesize fake samples ($x'$) and injects them into the learning process. By observing the responses of the honest participant, the adversary adjusts \( G \) to produce reconstructions that resemble the original training data. The adversarial objective is formulated as:
\begin{equation}
    A_{DRA} = \min_G \max_D \frac{1}{|X|} \sum_{x \in X} \log D(x) + \frac{1}{|X|} \sum_{x' \in X'} \log(1 - D(G(x')))
\end{equation}
where \( D \) is the discriminator. Through an iterative adversarial process, the adversary refines \( G \) to generate highly accurate reconstructions. Similarly, Zeng et al.~\cite{zeng2025gan} demonstrate how a semi-honest server in a USL setup can reconstruct private client data using GANs. The attack begins with a shadow model, where the server approximates the client’s model via an auxiliary dataset \( x_{{aux}}\). A GAN discriminator, combined with cross-entropy loss, ensures the shadow model's output aligns with $z_c$. Once trained, an inverse model maps features back to the input space, optimizing data reconstruction with MSE loss. While this exploits server-side vulnerabilities, SL also faces threats from malicious clients in multi-client setups~\cite{zeng2025gan}. An honest but curious client can extract the server model via knowledge distillation, aligning an alternative model with the global server model by minimizing KL-Div loss. The client then performs feature space inversion, using GAN-based optimization to infer missing class samples. Instead of reconstructing data in pixel space, the attack operates in a low-dimensional noise space, improving efficiency and inference accuracy.
\paragraph{v. Feature Reconstruction}
Feature-oriented reconstruction attack~\cite{xu2024stealthy} exploits the representation preference encoded in the $z_c$ that the client transmits to the server during training by a semi-honest server in the VansL setup. The attack follows a three-phase pipeline: First, the adversary constructs a substitute client $\tilde{f}_c$ by minimizing the distance between the client's smashed data and the generated features using a domain discriminator (DISC) network~\cite{ganin2015unsupervised,goodfellow2014generative}, which distinguishes between features from different domains, and Multi-Kernel Maximum Mean Discrepancy (MK-MMD)~\cite{gretton2012optimal,long2015learning}, a statistical measure to align distributions, to align feature spaces formulated as:
$
    \min_{\tilde{f}_c} L_{DISC} + L_{MK-MMD}
$
Second, an inverse mapping network, $\tilde{f_c^{-1}}$, is trained on public auxiliary data to reconstruct inputs from smashed data. Finally, during the attack phase, the trained inverse network is applied to the victim’s smashed data snapshot $z_c$ to reconstruct the private training data, given by: $\tilde{x} = f_c^{-1}(z_c)$.

Zhu et. al~\cite{zhu2023passive} achieve feature reconstruction by employing a simulator model trained on an auxiliary dataset that follows a similar distribution as the client’s private data in a VanSL setup. This simulator aims to approximate the behavior of the client’s private model, enabling the semi-honest server to infer private data without direct access. To enhance its effectiveness, adversarial regularization is introduced through a discriminator network $D_1$ that distinguishes real intermediate representations from synthetic ones, ensuring that the simulator learns indistinguishable representations. Once trained, a decoder model is used to reconstruct the original input features from the intermediate representations, with an additional discriminator $D_2$ guiding it to produce reconstructions that closely resemble real data.
\section{Details of Label Inference Attacks} 
This section provides an in-depth examination of the label inference attacks discussed in Section~\ref{Label Inference Attacks}, elaborating on their mathematical foundations and implementation details.
\subsection{Details of Gradient-Based Label Inference:}\label{sec:labelinferenceExtended}
The ExPLoit framework~\cite{kariyappa2023exploit} %proposed by Kariyappa et al. 
formulates label inference as an optimization problem where the surrogate labels $\tilde{\hat{y}}$ are iteratively refined. The server initializes random surrogate labels and computes $L' = H(\tilde{\hat{y},\tilde{\hat{p}}})$ where $(\tilde{\hat{p}})$ are predictions from the surrogate model. Through backpropagation $\tilde{fc^*}$, the attack computes $\nabla{z_s}L'$ to minimize the ExPLoit loss function which consists of (1) gradient matching by minimizing \( \mathbb{E} [ \| \nabla_{z_s} L' - \nabla_{\tilde{z_s}} L \|^2 ] \), (2) label prior regularization via KL-Div \( D_{\text{KL}}(P_y \| P_{\hat{y}}) \), and (3) cross-entropy regularization \( \mathbb{E} [ H(\tilde{\hat{y}}, \tilde{\hat{p}}) / H(P_y) ] \), ensuring label alignment. The final loss function is:
\begin{equation}
    L_{\text{ExPL}} = \mathbb{E} \left[ \| \nabla_{z_s} L - \nabla_{\tilde{z_s}} L' \|^2 \right] + \lambda_{\text{ce}} \cdot \mathbb{E} \left[ \frac{H(\tilde{\hat{y}},\tilde{\hat{p}})}{H(P_y)} \right] + \lambda_p \cdot D_{\text{KL}}(P_y \| P_{\tilde{\hat{y}}})
\end{equation}
where \( \lambda_{\text{ce}} \) and \( \lambda_p \) are hyperparameters controlling the trade-off between objectives. By iteratively optimizing \( \tilde{f_c^*} \) and the inferred labels, the attack successfully recovers private labels.

Erdogan et al.~\cite{erdougan2022unsplit} attack methodology leverages gradient updates $\nabla_{z_c}L$ to achieve perfect label recovery accuracy by a semi-honest server. In their VanSL setup, the server receives $z_c$ and $\nabla_{z_c}L$ from the client. The attack randomly initializes $\tilde{f_c^*}$ matching the client model's architecture and iteratively tests all possible labels $\tilde{y}$, selecting the one that minimizes gradient differences:
\begin{equation}
\tilde{y}^* = \arg\min_{\tilde{y}} \text{MSE} \left( \frac{\partial L(f_c(f_s(x)), y)}{\partial \theta}, \frac{\partial L(\tilde{f_c^*}(f_s(x)), \tilde{y})}{\partial \tilde{\theta}} \right)
\end{equation}
This optimization approach computes the Mean Squared Error between the gradients of the original model parameters $(\theta)$ and those of the surrogate model parameters $(\tilde{\theta})$. The attack succeeds due to the deterministic relationship between labels and gradient patterns.

The embedding swapping attack~\cite{bai2023villain} operates through a structured process of gradient comparison. For implementation, the attacker first computes embeddings $z_t$ and $z_i$ for target and unknown samples respectively. During training, when \( x_i \) is used in a batch, the attacker first uploads \( z_i \) and records the corresponding back-propagated gradient \( \nabla{z_i} \) from the server. In a later batch, the attacker swaps the embedding by sending \( z_t \) instead and observes the new gradient update \( \tilde{\nabla{z_i}} \). The attacker determines whether \( x_i \) belongs to the target label based on the gradient norm ratio:
\begin{equation}
    \frac{\|\tilde{\nabla{z_i}}\|2}{\|\nabla{z_i}\|2} \leq \alpha
\end{equation}
and ensures that the gradient norm satisfies:
\begin{equation}
    \|\nabla{z_i}\|_2 \leq \mu
\end{equation}
where \( \|\cdot\|_2 \) represents the L2 norm, and \( \alpha, \mu \) are predefined threshold parameters. If both conditions hold, it suggests that swapping the embedding does not significantly impact the training loss, indicating that \( x_i \) likely belongs to the target label. This attack can be extended to multi-attacker cases.

Another attack proposed by Zhao et al.~\cite{zhao2024splitaum} enables a malicious client to infer private labels in VanSL through a structured three-step approach: dummy label initialization, auxiliary model training, and private label inference. The client first initializes an auxiliary model using semi-supervised clustering (K-Means) to generate structured dummy labels approximating the server’s label distribution. A small set of labeled auxiliary samples serves as cluster centroids, improving training efficiency. The auxiliary model is then trained with a composite loss function, aligning its predictions with the server’s outputs. Once trained, the model captures the server’s decision boundaries, allowing the client to infer private labels by forwarding inputs through both models. The auxiliary model training process involves optimizing a composite loss function comprising three components: 
\begin{itemize}
    \item \textbf{Distance-based loss} ($\ell_d$): Aligns gradients between the server and auxiliary model.
    \item \textbf{Performance-based loss} ($\ell_p$): Ensures predictions match dummy labels.
    \item \textbf{Knowledge-based loss} ($\ell_k$): Refines predictions using a small set of labeled auxiliary samples.
\end{itemize}
The iterative optimization process enables the auxiliary model to approximate the server’s classification behavior without direct access to labels. Once trained, the auxiliary model allows the client to infer private labels by forwarding inputs through both models:
\begin{equation}
\tilde{y} \leftarrow \tilde{f}(z_c)
\end{equation}
where $\tilde{y}$ represents the inferred labels, $\tilde{f}$ is the trained auxiliary model, and $f(X)$ is the client model’s intermediate representation.

Xie et. al~\cite{xie2023label} propose a label inference attack against VanSL under a regression setting, addressing a gap in prior research focused on classification tasks against a semi-honest server.
Their methodology consists of multiple stages. First, the attacker initializes dummy labels and a surrogate model that attempts to approximate the label model’s behavior. The attack then employs a gradient matching strategy, where the attacker iteratively updates the dummy labels to minimize the distance between the gradients of the surrogate model and the gradients received from the label party. To enhance attack effectiveness, the authors introduce learning regularization: (1) Gradient Distance Loss, which minimizes the discrepancy between the gradients of the surrogate and the original model, (2) Training Accuracy Loss, which ensures that the surrogate model’s predictions align with the dummy labels, and (3) Knowledge Learning Loss, which utilizes a small auxiliary dataset with known labels to guide the optimization process.

\subsection{Details of Smashed-Based Label Inference:}\label{Details of Smashed-Based Label Inference}
After the completion of training, the server no longer receives gradients, making gradient-based attacks ineffective. However, during the inference phase, the adversary still receives \( z_c \), which retains semantic similarities to \( x \), thereby encoding label information and enabling label inference. Liu et al.~\cite{liu2024similarity} introduced three primary approaches for smashed data inference: Euclidean Distance-Based Matching, Clustering-Based Inference, and Transfer Learning-Based Inference. The Euclidean distance approach assigns $z_c$ to the nearest stored reference sample $z_{ref}$, assuming they share the same label:
\begin{equation}
    j = \arg \min_{ref} \| z_c - z_{ref} \|^2
\end{equation}
The clustering-based method further groups similar smashed data into clusters using K-Means clustering. Additionally, the Transfer Learning-Based Inference approach leverages pre-trained models to extract feature representations from the smashed data, improving label inference accuracy even when the cut layer is far from the output. Specifically, the adversary applies a pre-trained model $f_\text{pretrained}$ to transform $z_c$ into a feature space where class separability is enhanced:
\begin{equation} 
    z_{\text{trans}} = f_{\text{pretrained}}(z_c) 
\end{equation}
Then, label inference is performed by matching $z_\text{trans}$ to the closest reference embedding $z_\text{ref,trans}$ in the feature space:
\begin{equation} 
    j = \arg \min_{ref} | z_{\text{trans}} - z_{\text{ref, trans}} |^2
\end{equation}
By leveraging knowledge from pre-trained models, the adversary can generalize across datasets and improve inference robustness. 

Extending beyond feature reconstruction,~\cite{zhu2023passive} also performs label inference in USL, where the client retains both the first and last layers of the model, keeping labels private. In this scenario, the server additionally trains a label simulator $\tilde{h}$ to approximate the client’s final model layers. Instead of directly learning label mappings, $\tilde{h}$ processes the received $z_{cr}$ and outputs $\hat{y}$. A key challenge in this setting is overfitting, if $\tilde{h}$ is trained solely on an auxiliary data $x_{aux}$, it may struggle to generalize to $x$. To address this, a random label-flipping mechanism is introduced, where a fraction of training labels are intentionally perturbed. This forces $\tilde{h}$ to learn generalized representations rather than memorizing specific label distributions from $x_{aux}$.

Huang et al.~\cite{huang2023pixel} proposed an attack on SFL that reconstructs pixel-wise accurate private training data from shared smashed data under a semi-honest server threat model. The attack comprises two phases: training and inference. During training, the server collects $z_c$ from honest clients and generates pseudo-samples $x'$ using a pseudo-sample generator trained with one-hot and entropy losses to ensure class-balanced generation. The model is then trained to learn the inverse mapping from smashed data to private images. In the inference phase, the model reconstructs private images from new smashed data. To enhance the effectiveness of the reconstruction, the model is optimized using cycle-consistency losses. The forward cycle consistency loss ensures that the reconstructed images, when processed again through the client-side model, produce outputs similar to the original smashed data. The backward cycle consistency loss ensures that processed smashed data can regenerate original-like images when passed through the model. The forward cycle-consistency loss is formally defined as:
\begin{equation}
L_f = \| \tilde{f}^{-1}(z_c) - x' \|^2
\end{equation}
These loss functions refine the inversion process, enabling more precise reconstructions of private data.

\section{Details of Model Manipulation Attacks}\label{sec:Model ManipulationExtended}
This section provides an in-depth examination of the model manipulation attacks discussed in Section~\ref{Model Manipulation Attacks}, elaborating on their mathematical foundations and implementation details.
\subsection{Details of Adversarial Attacks:}\label{sec:Adversarial Attacks Extended}
The adversary in~\cite{fan2023robustness} perturbs feature representations by minimizing their cosine similarity. The perturbation is iteratively updated as:
\begin{equation}
\delta = \text{Clip}_{\epsilon} \left\{ \delta - \zeta \frac{\partial L_{\text{attack}}(z_s, z_s^*)}{\partial \delta} \right\}
\end{equation}
where \( \zeta \) is the update step size, \( \epsilon \) is the perturbation budget, and \( \text{Clip}_{\epsilon}\{\cdot\} \) ensures the perturbation remains within the allowed \( \epsilon \)-ball.

The adversarial example in~\cite{he2024advusl} \( \tilde{x} \) is iteratively updated using a gradient-based method while ensuring that the perturbation \( \delta \) remains within the predefined budget:
\begin{equation}
\|\delta\|_{\infty} \leq \epsilon
\end{equation}
where \( \delta \) is the adversarial perturbation applied to the input \( x \), and \( \epsilon \) defines the perturbation constraint.

\subsection{Details of Backdoor and Poisoning Attack} \label{sec:Backdoor and Poisoning Attack Extended}
\par\textit{i. Label Flipping Attacks:}
%Recent research has identified several sophisticated attack methodologies targeting SL systems. 
Kohankhaki et al.~\cite{kohankhaki2023detecting} investigated static label flipping attacks, where malicious clients systematically modify class labels within their local training data before transmission to the server. These modifications remain constant throughout the training process, causing the VanSL model to develop incorrect associations that degrade its performance through increased misclassification rates and reduced generalization capability. Their study examined poisoning scenarios by varying the number of malicious clients (\( M \in \{0, 2, 4, 6, 8, 10\} \)) and poisoning rate per client (\( p \in \{0.25, 0.5, 0.75\} \)). The attacks specifically targeted ECG readings, flipping labels between normal and abnormal classes, thereby introducing systematic errors in the model's ability to distinguish between normal and abnormal readings.
\par\textit{ii. Client-Side Backdoor Attacks:}
Yu et. al~\cite{yu2023backdoor} introduce client-side backdoor attacks where a malicious client in the VanSL setup leverages its control over local training data to inject backdoor samples by modifying features or labels. If the client holds both the features and the labels, it directly backdoors the data by adding a trigger pattern and associating it with an incorrect label. If the server holds the labels, the attacker may use label inference techniques to identify and manipulate specific training samples. To enhance attack persistence, an auxiliary model is introduced to distinguish between the clean and backdoor samples in the feature space, improving the sensitivity of the model to backdoor patterns. This ensures that the attack remains effective without degrading the model’s performance on the primary task, making detection challenging.

\begin{figure}[htbp]
    \centering    
    \includegraphics[width=0.75\linewidth]{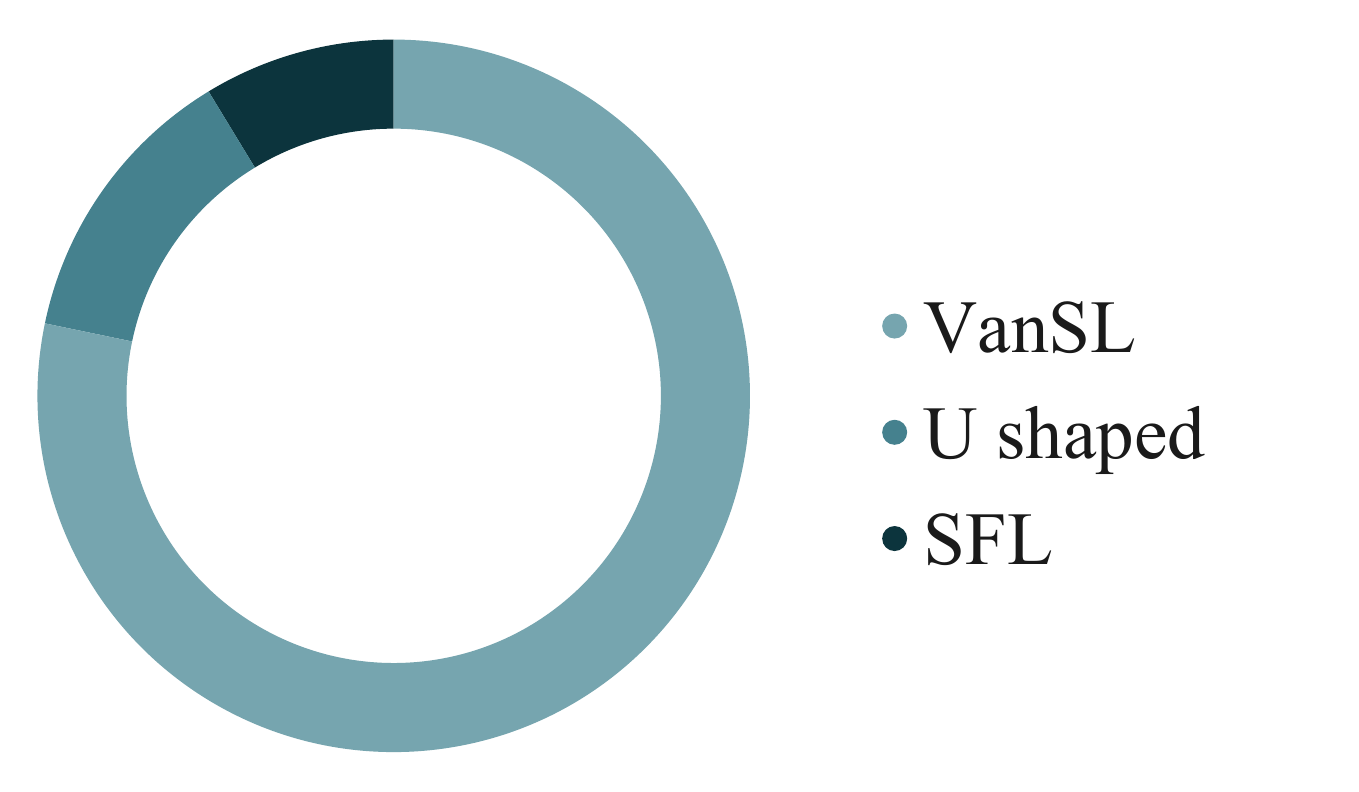}
    \caption{The distribution of split learning types in the surveyed literature on defense techniques.}
%    \Description{A diagram that showing distribution of split learning styles.}
    \label{fig:Split learning type pie}
\end{figure}
\begin{figure}[htbp]
  \centering  \includegraphics[width=0.35\textwidth]{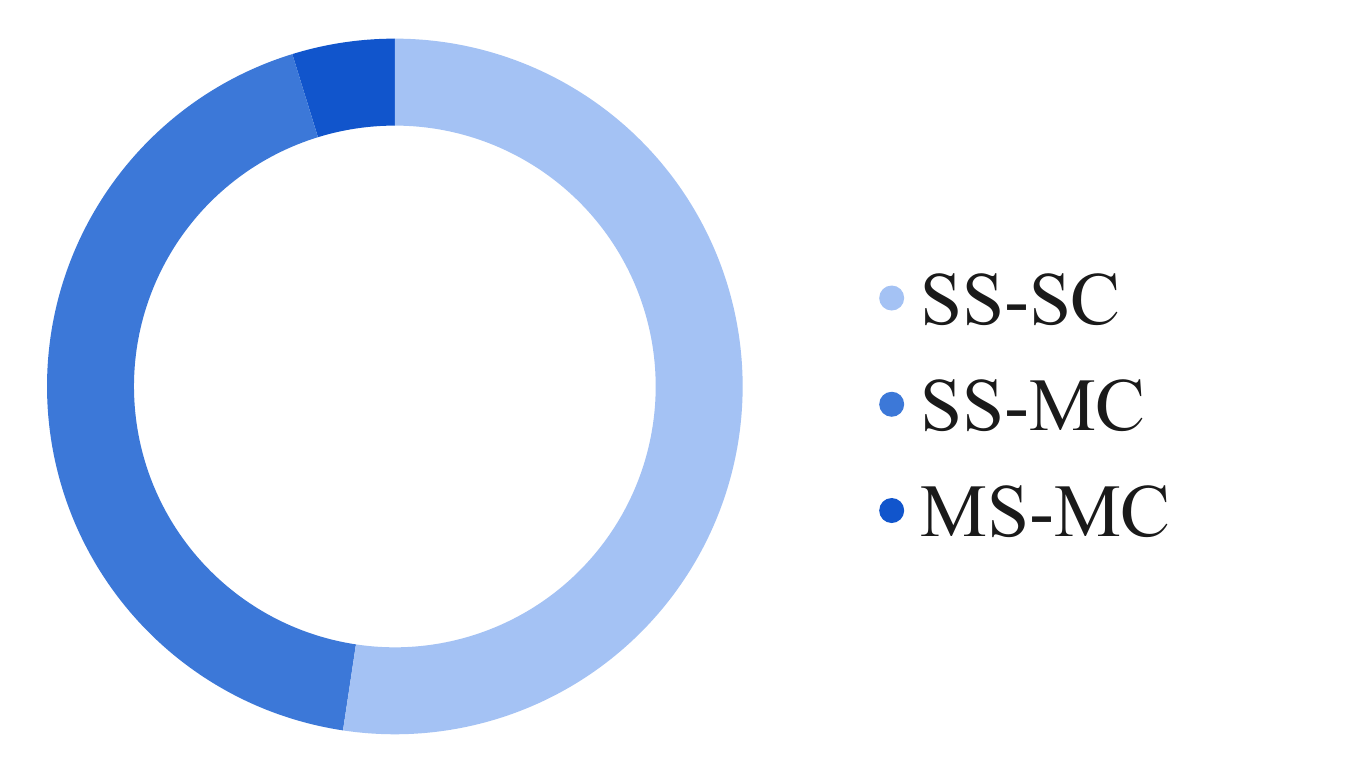} \caption{The distribution of client-server architectures in the surveyed literature on defense mechanisms.}
%  \Description{A diagram that showing distribution of client-server architectures.}
  \label{fig:Client Server distribution}
\end{figure}
Rieger et. al~\cite{safeSplit2025rieger} also examine client-side backdoor attacks in the USL setup. The attack process begins with trigger insertion into a subset of its training data. The malicious client then trains its local model on both clean and poisoned data, ensuring that the backdoor is embedded while maintaining high overall model accuracy. Once the adversarial client completes training, it transmits the model update to the server, which forwards it to the next client. This gradual backdoor propagation allows the backdoor to persist through multiple training rounds, as subsequent benign clients unknowingly build upon the manipulated model. Since SL does not reset model updates, the backdoor remains embedded even after multiple training rounds.
\par\textit{iii. Server-Side Backdoor Attacks:}
Further advancing the field of backdoor attacks, Tajalli et al.~\cite{tajalli2023feasibility} investigated server-side backdoor attacks in VanSL, where an adversarial server injects backdoors without direct access to the client data. They proposed two strategies: Surrogate Client Attack and Injector Autoencoder Attack. In the Surrogate Client Attack, the server introduces a surrogate client \( \hat{f}_c^* \) mimicking \( f_c \) and trains it on a poisoned dataset using a weighted loss function:$L_{\text{comb}} = \alpha L_c + (1 - \alpha) \tilde{L}_c^*$
where \( \alpha \) controls client influence in backpropagation. The Injector Autoencoder Attack trains an autoencoder on paired (clean, poisoned) smashed data and places it between the client's cut layer and the server’s input pipeline to modify incoming activations in real-time. Their results suggest VanSL is resilient to backdoor attacks. Contrarily, Yu et al.~\cite{yu2023backdoor} demonstrated that a malicious server can implant backdoors via feature space hijacking, aligning the client model’s optimization process with a shadow model trained on a separate dataset. A discriminator network transfers the backdoor’s effect to the client, while an auxiliary model maintains feature separability, ensuring stealth and preserving task accuracy. A discriminator network aids in transferring the backdoor’s effect to the client model, ensuring it learns to associate trigger patterns with attacker-specified outputs. An auxiliary model further reinforces backdoor persistence by maintaining feature separability. Since this method does not alter raw client data, it remains stealthy while ensuring the primary task’s accuracy is unaffected.

Recently, Yu et al.~\cite{yu2024chronic} introduced the SFI (Steal, Finetune, and Implant) attack framework, a sophisticated server-side backdoor attack in a VanSL framework, manipulating gradient updates without requiring access to raw client data. The framework operates in three distinct stages. In the Steal Stage, it constructs a shadow model $\tilde{f}_s$ using a limited shadow dataset $x_s$, labeling samples as either backdoor ($x_{sb} = 1$) or clean ($x_{sb} = 0$), while the shadow model learns from the server's main task model $f_s$. During the Finetune Stage, the framework optimizes the server model $f_s$, shadow model $\tilde{f}_s$, and auxiliary model $f_a$, where the auxiliary model assists in backdoor sample differentiation, ensuring stable backdoor encoding while maintaining main task accuracy. Finally, in the Implant Stage, the framework transfers backdoor capability to the client model by implementing a discriminator for adversarial training, forcing the client model to adapt feature encoding to match the shadow model while maintaining original training protocol integrity.

\section{Supplementary Tables and Figures} \label{sec:supplementaryTable}
We provide the general overview of attack and defense strategies and their features, such as implementation availability and SL style in Table~\ref{tab:general table}. We also present the distribution of defense techniques across multi-client and multi-server settings in Figure~\ref{fig:Client Server distribution} and the distribution of these SL types across the reviewed literature in Figure~\ref{fig:Split learning type pie}.
\input{generalTable}

%% file: generalTable.tex
\begin{table*}[ht]
% \small
\centering
\begin{adjustbox}{width=0.95\textwidth} 
% \begin{tabular}
\setlength{\tabcolsep}{4pt}
\begin{tabular}{ccccccc}

\toprule
\textbf{Paper} & \textbf{Split Learning Style} & \textbf{Implementation} & \textbf{Training Setup} & \textbf{Model Architecture}  &\textbf{Attack Strategy}& \textbf{Defense Strategy} \\ \midrule

Fu et al.\cite{Pinocchio} &
SFL &
\href{https://github.com/CGCL-codes/GradientsScrutinizer}{Available} &
SS-SC &
DNN & \multicolumn{1}{c}{NA}&
Monitoring - Gradient Detection
\\ \midrule

Erdogan et al.\cite{erdogan2023defense} &
VanSL &
\href{https://github.com/ege-erdogan/splitguard}{Available} &
SS-SC &
DNN (ResNet)  & \multicolumn{1}{c}{NA}&
Monitoring - Gradient Detection \\ \midrule

SplitOut\cite{erdogan2024splitout} &
VanSL &
\href{https://github.com/ege-erdogan/splitout}{Available} &
SS-SC &
DNN (ResNet)  & \multicolumn{1}{c}{NA}&
Monitoring - Gradient Detection
\\ \midrule

Titcombe et al.\cite{practical2021} &
VanSL &
\href{https://github.com/TTitcombe/Model-Inversion-SplitNN}{Available} &
SS-MC &
DNN  & \multicolumn{1}{c}{NA}&
Data Perturbation
\\ \midrule

% Fusion &
% SFL &
% \href{https://github.com/daisy611/Fusion}{Available} &
% SS-MC &
% DNN &
% Monitoring - Weight/anomaly detection
% \\ \midrule

Mao et al. \cite{mao2023securesplit} &
VanSL &
No Open Source &
SS-SC &
DNN  & \multicolumn{1}{c}{NA}&
Architecture Modification - Data Decorrelation
\\ \midrule

Khan et al. \cite{khan2024make} &
VanSL &
\href{https://github.com/UnoriginalOrigi/SplitFSS}{Available} &
MS-SC &
DNN  & \multicolumn{1}{c}{NA}&
Secure Computation - Function Secret Sharing
\\ \midrule

Khowaja et al. \cite{khowaja2024foesfooled} &
VanSL &
No Open Source &
SS-SC &
DNN & \multicolumn{1}{c}{NA}&
Architecture Modification - Data Perturbation
\\ \midrule

ResSFL \cite{Li2022ResSFL} &
SFL &
\href{https://github.com/zlijingtao/ResSFL}{Available} &
SS-MC &
DNN (VGG-11)  & \multicolumn{1}{c}{NA}&
Architecture Modification - Protocol Modification
\\ \midrule

% DLG\cite{zhu2019dlg} &
% Not Specified &
% No Open Source &
% MC-SS &
% DNN &
% Gradient Perturbation, Noise Injection 
% \\ \midrule

Pham et al. \cite{pham2024splitlearninglocalweight} &
VanSL &
No Open Source &
MS-MC &
DNN  & \multicolumn{1}{c}{NA}&
Architecture Modification - Protocol Modification \\ \midrule

PrivateMail \cite{PrivateMail} &
VanSL &
No Open Source &
SS-SC &
DNN  & \multicolumn{1}{c}{NA}&
Data Perturbation - Differential Privacy
\\ \midrule

% % Shredder\cite{Shredder} &
% % VanSL &
% % No Open Source &
% % SS-SS &
% % DNN &
% % Differential Privacy
% % \\ \midrule

% HashVFL\cite{qiu2022hashvfl} &
% VSL &
% No Open Source &
% MC &
% DNN &
% Differential Privacy, Gradient Privacy
% \\ \midrule

Pham et al. \cite{pham2022binarizing} &
VanSL &
\href{https://github.com/phamngocduy/BinarizeLocalizedLayers}{Available} &
SS-MC &
DNN  & \multicolumn{1}{c}{NA}&
Architecture Modification \& Differential Privacy \\ \midrule

DISCO\cite{DISCO} &
VanSL &
\href{https://github.com/aidecentralized/InferenceBenchmark}{Available} &
SS-SC &
DNN  & \multicolumn{1}{c}{NA}&
Architecture Modification - Data Decorrelation
\\ \midrule

Turina et al. \cite{turina2021fsl} &
SFL & 
No Open Source &
SS-MC &
DNN  & \multicolumn{1}{c}{NA}&
Architecture Modification - Data Decorrelation\\ \midrule

% NoPeek\cite{NoPeek} & 
% VanSL & 
% No Open Source & 
% SS-SC & 
% DNN (ResNet-18) & 
% Architecture Modification - Data Decorrelation 
% \\ \midrule

% NoPeek-Infer \cite{nopeek_infer_2021} &
% VanSL &
% No Open Source &
% SS-SS &
% DNN (ResNet-18) &
% Distance Correlation 
% \\ \midrule

Gawron et al. \cite{gawron2022feature} &
VanSL &
\href{https://github.com/pasquini-dario/SplitNN_FSHA}{Available} &
SS-SC &
DNN  & \multicolumn{1}{c}{NA}&
Differential Privacy\\ \midrule

Khan et al. \cite{LoveHate} &
USL &
\href{https://github.com/khoaguin/HESplitNet}{Available} &
SS-SC &
DNN (1D CNN)  & \multicolumn{1}{c}{NA}&
Secure Computation - Homomorphic Encryption
\\ \midrule

Khan et al. \cite{khan2023more_secure_split} &
USL &
\href{https://github.com/khoaguin/HESplitNet}{Available} &
SS-SC &
1D CNN  & \multicolumn{1}{c}{NA}&
Secure Computation - Homomorphic Encryption
\\ \midrule

Split HE\cite{splitHE} &
VanSL &
No Open Source &
SS-SC &
DNN  & \multicolumn{1}{c}{NA}&
Secure Computation - Homomorphic Encryption
\\ \midrule

Split Ways\cite{SplitWays2023} &
USL &
No Open Source &
SS-SC &
1D CNN  & \multicolumn{1}{c}{NA}&
Secure Computation - Homomorphic Encryption
\\ \midrule

Nguyen et al. \cite{SplitWithoutALeak} &
USL &
\href{https://github.com/khoaguin/HESplitNet}{Available} &
SS-SC &
1D CNN  & \multicolumn{1}{c}{NA}&
Secure Computation - Homomorphic Encryption
\\ \midrule

Abuadbba et al.\cite{Abuadbba2020} &
VanSL &
\href{https://github.com/SharifAbuadbba/split-learning-1D}{Available} &
SS-MC &
1D CNN  & \multicolumn{1}{c}{NA}&
Architecture Modification \& Differential Privacy \\ \midrule

% Ryu et al.\cite{Ryu2023} &
% VanSL &
% No Open Source &
% MC-SS &
% DNN &
% Differential Privacy
% \\ \midrule

SafeSplit\cite{safeSplit2025rieger} &
USL &
No Open Source &
SS-MC &
DNN  & \multicolumn{1}{c}{NA}&
Monitoring - Weight/anomaly detection
\\ \midrule

PSLF\cite{PSLF2023Wan} &
VanSL &
No Open Source&
SS-SC &
DNN  & \multicolumn{1}{c}{NA}&
Architecture Modification - Data Decorrelation
\\ \midrule

% Analyze SplitFed\cite{AnalyzeSplitFed2023Ismail} &
% HSL &
% No Open Source &
% MC-SS &
% DNN, 1D-CNN &
% None \\ \midrule

% Optimized secure CNN\cite{OptimizedCNN2023Kim} &
% HSL &
% No Open Source &
% MC-SS &
% DNN &
% Homomorphic Encryption \\ \midrule

CURE\cite{CURE2024Kanpak} &
VanSL &
\href{https://github.com/hkanpak21/CURE}{Available} &
SS-SC &
DNN  & \multicolumn{1}{c}{NA}&
Secure Computation - Homomorphic Encryption
\\ \arrayrulecolor{black!50}\specialrule{3pt}{2\jot}{1pc}

Pasquini et al. \cite{Unleashing_tiger} &
VSL &
\href{https://github.com/pasquini-dario/SplitNN_FSHA}{Available} &
SS-MC &
DNN  &
Data reconstruction - FSHA&
 \multicolumn{1}{c}{NA}  \\ \midrule

PCAT \cite{zhang2024functionality} &
VanSL &
No Open Source &
SS-SC &
CNN  &
Data reconstruction - Functionality Stealing&
 \multicolumn{1}{c}{NA} 
\\ \midrule

Unsplit\cite{erdougan2022unsplit} &
VanSL &
\href{https://github.com/ege-erdogan/unsplit.}{Available} &
SS-SC &
DNN  &
Data Reconstruction - Model Inversion&
 \multicolumn{1}{c}{NA} 
\\ \midrule

Xu et al. \cite{xu2024stealthy} &
VanSL &
No Open Source &
SS-SC &
DNN  &
Data Reconstruction - Feature Reconstruction&
 \multicolumn{1}{c}{NA} 
\\ \midrule

Li et al.\cite{li2021label} &
VanSL &
No Open Source &
SS-SC &
DNN  &
Label Inference - Gradient-Based Label Inference&
 \multicolumn{1}{c}{NA} 
\\ \midrule

Liu et al.\cite{liu2024similarity} &
VanSL &
\href{https://github.com/ZeroWalker10/sl_similarity_label_inference.}{Available} &
SS-SC &
DNN  &
Label Inference - Gradient-Based Label Inference&
 \multicolumn{1}{c}{NA} 
\\ \midrule

ExPLoit\cite{kariyappa2023exploit} &
VanSL &
No Open Source &
SS-SC &
DNN  &
Label Inference - Gradient-Based Label Inference&
 \multicolumn{1}{c}{NA} 
\\ \midrule

VILLIAN\cite{bai2023villain} &
VSL &
No Open Source &
SS-SC &
DNN  &
Model Manipulation - Backdoor Attacks&
 \multicolumn{1}{c}{NA} 
\\ \midrule

Yu et al. \cite{yu2024chronic} &
VanSL &
No Open Source &
SS-SC &
DNN  &
Model Manipulation - Backdoor Attacks&
 \multicolumn{1}{c}{NA} 
\\ \midrule

Tajalli et al.\cite{tajalli2023feasibility} &
VanSL &
No Open Source &
SS-SC &
DNN  &
Model Manipulation - Backdoor Attacks&
 \multicolumn{1}{c}{NA} 
\\ \midrule

Fan et al.\cite{fan2023robustness} &
USL &
No Open Source &
SS-SC &
DNN  &
Model Manipulation - Adversarial Attacks&
 \multicolumn{1}{c}{NA} 
\\ \midrule

AdvUSL\cite{he2024advusl} &
USL &
No Open Source &
SS-MC &
DNN  &
Model Manipulation - Adversarial Attacks&
 \multicolumn{1}{c}{NA} 
\\ \midrule

Kohankhaki et al.\cite{kohankhaki2023detecting} &
VanSL &
\href{ https://github.com/a-ayad/Split-ECG-Classification.}{Available} &
SS-MC &
DNN  &
Model Manipulation - Poisoning Attacks&
 \multicolumn{1}{c}{NA} 
\\ \midrule

Zhu et al.\cite{zhu2023passive} &
USL &
No Open Source &
SS-MC &
DNN  &
Data Reconstruction- Feature Reconstruction&
 \multicolumn{1}{c}{NA} 
\\ \midrule

Yu et al.\cite{yu2023backdoor} &
VanSL &
No Open Source &
SS-MC &
DNN  &
Model Manipulation - Backdoor Attacks&
 \multicolumn{1}{c}{NA} 
\\ \midrule

SplitAum\cite{zhao2024splitaum} &
VanSL &
No Open Source &
SS-SC &
DNN  &
Label Inference - Gradient Based Label Inference&
 \multicolumn{1}{c}{NA} 
\\ \midrule

Zeng et al.\cite{zeng2025gan} &
USL &
No Open Source &
SS-SC &
DNN  &
Data reconstruction - GAN&
 \multicolumn{1}{c}{NA} 
\\ \midrule

Huang et al.\cite{huang2023pixel} &
SFL &
No Open Source &
MS-MC &
DNN  &
Label Inference - Smashed Based Label Inference&
 \multicolumn{1}{c}{NA} 
\\ \midrule

Gajbhiye et al.\cite{gajbhiye2022data} &
SFL &
No Open Source &
MS-MC &
DNN  &
Model Manipulation - Poisoning Attack&
 \multicolumn{1}{c}{NA} 
\\ \midrule

Xie et al.\cite{xie2023label} &
VanSL &
\href{https://github.com/xiehahha/aaai_ppai23_split_learning_leakage}{Available} &
SS-SC &
DNN  &
Label Inference - Gradient Based Label Inference&
 \multicolumn{1}{c}{NA} 
\\ \midrule

Wu et al.\cite{wu2024evaluating} &
SFL &
\href{https://github.com/xxxcuss/asdf456jkk}{Available} &
MS-MC &
DNN  &
Model Manipulation - Poisoning Attack&
 \multicolumn{1}{c}{NA} 
\\ \midrule

Ismail et al.\cite{AnalyzeSplitFed2023Ismail} &
SFL &
No Open Source &
MS-MC &
DNN  &
Model Manipulation - Poisoning Attack&
 \multicolumn{1}{c}{NA} 
\\ \bottomrule
\\
\end{tabular}
\end{adjustbox}
\caption{Comparison of prior Split Learning approaches, showing key attributes such as \emph{VanSL} (Vanilla SL), \emph{SFL} (Split Federated Learning), \emph{USL} (U-Shaped SL), \emph{VSL} (Vertical SL), \emph{SS-SC} (Single-Server Single-Client), \emph{SS-MC} (Single-Server Multi-Client), \emph{MS-SC} (Multi-Server Single-Client), and \emph{MS-MC} (Multi-Server Multi-Client). Model abbreviations include \emph{DNN} (Dense Neural Network), \emph{CNN} (Convolutional Neural Network) and \emph{1D CNN} (1 Dimensional Convolutional Neural Network). The thick gray line separates \emph{defense} papers (top) from \emph{attack} papers (bottom) and NA stands for 'not applicable'.}
% \label{split_learning_architectures}
\label{tab:general table}
\end{table*}

%% file: elsarticle-template-num-names.bbl
\begin{thebibliography}{10}

\bibitem{Abuadbba2020}
S.~Abuadbba, K.~Kim, M.~Kim, C.~Thapa, S.~A. Camtepe, Y.~Gao, H.~Kim, and S.~Nepal.
\newblock Can we use split learning on 1d cnn models for privacy preserving training?
\newblock In {\em ASIACCS}, 2020.

\bibitem{arora2012online}
R.~Arora, O.~Dekel, and A.~Tewari.
\newblock Online bandit learning against an adaptive adversary: from regret to policy regret.
\newblock {\em arXiv:1206.6400}, 2012.

\bibitem{aumann2010security}
Y.~Aumann and Y.~Lindell.
\newblock Security against covert adversaries: Efficient protocols for realistic adversaries.
\newblock {\em Journal of Cryptology}, 2010.

\bibitem{bai2023villain}
Y.~Bai, Y.~Chen, H.~Zhang, W.~Xu, H.~Weng, and D.~Goodman.
\newblock $\{$VILLAIN$\}$: Backdoor attacks against vertical split learning.
\newblock In {\em USENIX Security}, 2023.

\bibitem{boyle2015function}
E.~Boyle, N.~Gilboa, and Y.~Ishai.
\newblock Function secret sharing.
\newblock In {\em EUROCRYPT}. Springer, 2015.

\bibitem{carlini2021encoding}
N.~Carlini, S.~Deng, S.~Garg, S.~Jha, S.~Mahloujifar, M.~Mahmoody, S.~Song, A.~Thakurta, and F.~Tram{\`e}r.
\newblock Is private learning possible with instance encoding?
\newblock 2021.

\bibitem{Chen2024}
Z.~Chen, H.~Zheng, and G.~Liu.
\newblock Awdp-fl: An adaptive differential privacy federated learning framework.
\newblock {\em Electronics}, 2024.

\bibitem{debenedetti_sidechannel}
E.~Debenedetti, G.~Severi, N.~Carlini, C.~A. Choquette-Choo, M.~Jagielski, M.~Nasr, E.~Wallace, and F.~Tramèr.
\newblock Privacy side channels in machine learning systems.
\newblock In {\em USENIX Security}, 2024.

\bibitem{dwork2006differential}
C.~Dwork.
\newblock Differential privacy.
\newblock In {\em ICALP}, 2006.

\bibitem{erdougan2022unsplit}
E.~Erdo{\u{g}}an, A.~K{\"u}p{\c{c}}{\"u}, and A.~E. {\c{C}}i{\c{c}}ek.
\newblock Unsplit: Data-oblivious model inversion, model stealing, and label inference attacks against split learning.
\newblock In {\em ACM WPES}, 2022.

\bibitem{erdogan2024splitout}
E.~Erdogan, U.~Teksen, M.~S. Celiktenyildiz, A.~Kupcu, and A.~E. Cicek.
\newblock Splitout: Out-of-the-box training-hijacking detection in split learning via outlier detection.
\newblock In {\em CANS}, 2024.

\bibitem{erdogan2022splitguard}
E.~Erdoğan, A.~Küpçü, and A.~E. Çiçek.
\newblock Splitguard: Detecting and mitigating training-hijacking attacks in split learning.
\newblock In {\em WPES}, 2022.

\bibitem{erdogan2023defense}
E.~Erdoğan, U.~Tekşen, M.~S. Çeliktenyıldız, A.~Küpçü, and A.~E. Çiçek.
\newblock Defense mechanisms against training-hijacking attacks in split learning.
\newblock {\em IEEE TKDE}, 2023.

\bibitem{fan2023robustness}
M.~Fan, C.~Chen, C.~Wang, W.~Zhou, and J.~Huang.
\newblock On the robustness of split learning against adversarial attacks.
\newblock In {\em ECAI}. 2023.

\bibitem{fang2020local}
M.~Fang, X.~Cao, J.~Jia, and N.~Gong.
\newblock Local model poisoning attacks to $\{$Byzantine-Robust$\}$ federated learning.
\newblock In {\em USENIX Security}, 2020.

\bibitem{fu2022label}
C.~Fu, J.~Zhang, T.~Zhu, W.~Zhou, and P.~S. Yu.
\newblock Label inference attacks against vertical federated learning.
\newblock In {\em USENIX Security}, 2022.

\bibitem{fu2022adap}
J.~Fu, Z.~Chen, and X.~Han.
\newblock Adap dp-fl: Differentially private federated learning with adaptive noise, 2022.

\bibitem{Pinocchio}
J.~Fu, X.~Ma, B.~B. Zhu, P.~Hu, R.~Zhao, Y.~Jia, P.~Xu, H.~Jin, and D.~Zhang.
\newblock Focusing on pinocchio's nose: A gradients scrutinizer to thwart split-learning hijacking attacks using intrinsic attributes.
\newblock In {\em NDSS}, 2023.

\bibitem{gajbhiye2022data}
S.~Gajbhiye, P.~Singh, and S.~Gupta.
\newblock Data poisoning attack by label flipping on splitfed learning.
\newblock In {\em RTIP2R}, 2022.

\bibitem{ganin2015unsupervised}
Y.~Ganin and V.~Lempitsky.
\newblock Unsupervised domain adaptation by backpropagation.
\newblock In {\em ICML}, 2015.

\bibitem{gao2022combined}
Y.~Gao, M.~Du, X.~Zhang, and Y.~Xiang.
\newblock Combined federated and split learning in edge computing: Taxonomy and open issues.
\newblock {\em Sensors}, 2022.

\bibitem{gawron2022feature}
G.~Gawron and P.~Stubbings.
\newblock Feature space hijacking attacks against differentially private split learning.
\newblock {\em arXiv:2201.04018}, 2022.

\bibitem{goodfellow2014generative}
I.~Goodfellow, J.~Pouget-Abadie, M.~Mirza, B.~Xu, D.~Warde-Farley, S.~Ozair, A.~Courville, and Y.~Bengio.
\newblock Generative adversarial nets.
\newblock {\em NIPS}, 2014.

\bibitem{gretton2012optimal}
A.~Gretton, D.~Sejdinovic, H.~Strathmann, S.~Balakrishnan, M.~Pontil, K.~Fukumizu, and B.~K. Sriperumbudur.
\newblock Optimal kernel choice for large-scale two-sample tests.
\newblock {\em NIPS}, 2012.

\bibitem{gupta2018split}
O.~Gupta and R.~Raskar.
\newblock Distributed learning of deep neural network over multiple agents, 2018.

\bibitem{he2024advusl}
Y.~He, C.~Hu, Y.~Pu, J.~Chen, and X.~Li.
\newblock Advusl: Targeted adversarial attack against u-shaped split learning.
\newblock In {\em IEEE MASS}, 2024.

\bibitem{he2019model}
Z.~He, T.~Zhang, and R.~B. Lee.
\newblock Model inversion attacks against collaborative inference.
\newblock In {\em ACSAC}, 2019.

\bibitem{Hitaj2017}
B.~Hitaj, G.~Ateniese, and F.~Perez-Cruz.
\newblock Deep models under the gan: Information leakage from collaborative deep learning.
\newblock In {\em ACM SIGSAC}, 2017.

\bibitem{hu2025review}
Z.~Hu, T.~Zhou, B.~Wu, C.~Chen, and Y.~Wang.
\newblock A review and experimental evaluation on split learning.
\newblock {\em Future Internet}, 2025.

\bibitem{huang2023pixel}
H.~Huang, X.~Li, and W.~He.
\newblock Pixel-wise reconstruction of private data in split federated learning.
\newblock In {\em ICICS}, 2023.

\bibitem{huang2024TEE}
W.~Huang, Y.~Wang, A.~Cheng, A.~Zhou, C.~Yu, and L.~Wang.
\newblock A fast, performant, secure distributed training framework for llm.
\newblock In {\em ICASSP}, 2024.

\bibitem{ismail2023analyzing}
A.~T.~Z. Ismail and R.~M. Shukla.
\newblock Analyzing the vulnerabilities in splitfed learning: Assessing the robustness against data poisoning attacks.
\newblock {\em arXiv:2307.03197}, 2023.

\bibitem{AnalyzeSplitFed2023Ismail}
A.~T.~Z. Ismail and R.~M. Shukla.
\newblock Analyzing the vulnerabilities in splitfed learning: Assessing the robustness against data poisoning attacks, 2023.

\bibitem{joshi2022performance}
P.~Joshi, C.~Thapa, S.~Camtepe, M.~Hasanuzzaman, T.~Scully, and H.~Afli.
\newblock Performance and information leakage in splitfed learning and multi-head split learning in healthcare data and beyond.
\newblock {\em Methods and Protocols}, 2022.

\bibitem{CURE2024Kanpak}
H.~I. Kanpak, A.~Shabbir, E.~Genç, A.~Küpçü, and S.~Sav.
\newblock Cure: Privacy-preserving split learning done right, 2024.

\bibitem{karakocc2024fault}
F.~Karako{\c{c}}, A.~K{\"u}p{\c{c}}{\"u}, and M.~{\"O}nen.
\newblock Fault tolerant and malicious secure federated learning.
\newblock In {\em CANS}, 2024.

\bibitem{kariyappa2023exploit}
S.~Kariyappa and M.~K. Qureshi.
\newblock Exploit: Extracting private labels in split learning.
\newblock In {\em SaTML}. IEEE, 2023.

\bibitem{khan2024make}
T.~Khan, M.~Budzys, and A.~Michalas.
\newblock Make split, not hijack: Preventing feature-space hijacking attacks in split learning.
\newblock In {\em SACMAT}, 2024.

\bibitem{khan2023more_secure_split}
T.~Khan, K.~Nguyen, and A.~Michalas.
\newblock A more secure split: Enhancing the security of privacy-preserving split learning.
\newblock In {\em AsiaCCS}. Tampere University, 2023.

\bibitem{SplitWays2023}
T.~Khan, K.~Nguyen, and A.~Michalas.
\newblock Split ways: Privacy-preserving training of encrypted data using split learning.
\newblock In {\em arXiv:2301.08778, 2023}, 2023.

\bibitem{LoveHate}
T.~Khan, K.~Nguyen, A.~Michalas, and A.~Bakas.
\newblock Love or hate? share or split? privacy-preserving training using split learning and homomorphic encryption, 2023.

\bibitem{khowaja2024foesfooled}
S.~A. Khowaja, I.~H. Lee, K.~Dev, M.~A. Jarwar, and N.~M.~F. Qureshi.
\newblock Get your foes fooled: Proximal gradient split learning for defense against model inversion attacks on iomt data.
\newblock {\em IEEE TNSE}, 2024.

\bibitem{kohankhaki2023detecting}
M.~Kohankhaki, A.~Ayad, M.~Barhoush, and A.~Schmeink.
\newblock Detecting data poisoning in split learning using intraclass-distance inflated loss.
\newblock In {\em IEEE GC Wkshps}, 2023.

\bibitem{Li2022ResSFL}
J.~Li, A.~S. Rakin, X.~Chen, Z.~He, D.~Fan, and C.~Chakrabarti.
\newblock Ressfl: A resistance transfer framework for defending model inversion attack in split federated learning.
\newblock In {\em CVPR}, 2022.

\bibitem{li2021label}
O.~Li, J.~Sun, X.~Yang, W.~Gao, H.~Zhang, J.~Xie, V.~Smith, and C.~Wang.
\newblock Label leakage and protection in two-party split learning.
\newblock {\em arXiv:2102.08504}, 2018.

\bibitem{li2022federated}
Z.~Li, S.~Si, J.~Wang, and J.~Xiao.
\newblock Federated split bert for heterogeneous text classification, 2022.

\bibitem{li2024split}
Z.~Li, C.~Yan, X.~Zhang, G.~Gharibi, Z.~Yin, X.~Jiang, and B.~A. Malin.
\newblock Split learning for distributed collaborative training of deep learning models in health informatics.
\newblock In {\em AMIA Annu. Symp. Proc}, 2024.

\bibitem{liu2024similarity}
J.~Liu, X.~Lyu, Q.~Cui, and X.~Tao.
\newblock Similarity-based label inference attack against training and inference of split learning.
\newblock {\em IEEE TIFS}, 2024.

\bibitem{long2015learning}
M.~Long, Y.~Cao, J.~Wang, and M.~Jordan.
\newblock Learning transferable features with deep adaptation networks.
\newblock In {\em PMLR}, 2015.

\bibitem{maeng2023fisher}
K.~Maeng, C.~Guo, S.~Kariyappa, and G.~E. Suh.
\newblock Bounding the invertibility of privacy-preserving instance encoding using fisher information.
\newblock {\em NeurIPS}, 2023.

\bibitem{mansouri2023sok}
M.~Mansouri, M.~{\"O}nen, W.~B. Jaballah, and M.~Conti.
\newblock Sok: Secure aggregation based on cryptographic schemes for federated learning.
\newblock {\em PoPETs}, 2023.

\bibitem{mao2023securesplit}
Y.~Mao, Z.~Xin, Z.~Li, J.~Hong, Q.~Yang, and S.~Zhong.
\newblock Secure split learning against property inference, data reconstruction, and feature space hijacking attacks.
\newblock In {\em ESORICS}, 2023.

\bibitem{nasr2018machine}
M.~Nasr, R.~Shokri, and A.~Houmansadr.
\newblock Machine learning with membership privacy using adversarial regularization.
\newblock In {\em ACM SIGSAC}, 2018.

\bibitem{Nasr2019Comprehensive}
M.~Nasr, R.~Shokri, and A.~Houmansadr.
\newblock Comprehensive privacy analysis of deep learning: Passive and active white-box inference attacks against centralized and federated learning.
\newblock In {\em IEEE S\&P}, 2019.

\bibitem{ng2023sok}
L.~K. Ng and S.~S. Chow.
\newblock Sok: cryptographic neural-network computation.
\newblock In {\em IEEE S\&P}, 2023.

\bibitem{pham2024e}
K.~T.~P. Ngoc Duy~Pham and N.~Chilamkurti.
\newblock Enhancing accuracy-privacy trade-off in differentially private split learning, 2024.

\bibitem{SplitWithoutALeak}
K.~Nguyen, T.~Khan, and A.~Michalas.
\newblock Split without a leak: Reducing privacy leakage in split learning.
\newblock In {\em SecureComm}, 2025.

\bibitem{openai2024gpt4o}
OpenAI.
\newblock Gpt-4o: Multimodal ai model.

\bibitem{Unleashing_tiger}
D.~Pasquini, G.~Ateniese, and M.~Bernaschi.
\newblock Unleashing the tiger: Inference attacks on split learning.
\newblock In {\em ACM CCS}, 2021.

\bibitem{peng2025survey}
Z.~Peng, T.~Wang, C.~Zhao, G.~Liao, Z.~Lin, Y.~Liu, B.~Cao, L.~Shi, Q.~Yang, and S.~Zhang.
\newblock A survey of zero-knowledge proof based verifiable machine learning, 2025.

\bibitem{splitHE}
G.-L. Pereteanu, A.~Alansary, and J.~Passerat-Palmbach.
\newblock Split he: Fast secure inference combining split learning and homomorphic encryption, 2022.

\bibitem{pham2022binarizing}
N.~D. Pham, A.~Abuadbba, Y.~Gao, T.~K. Phan, and N.~Chilamkurti.
\newblock Binarizing split learning for data privacy enhancement and computation reduction, 2022.

\bibitem{pham2023data}
N.~D. Pham and N.~Chilamkurti.
\newblock Data leakage threats and protection in split learning: A survey.
\newblock In {\em ICEA}, 2023.

\bibitem{Pham2024Enhancing}
N.~D. Pham, K.~T. Phan, and N.~Chilamkurti.
\newblock Enhancing accuracy-privacy trade-off in differentially private split learning.
\newblock {\em IEEE TIFS}, 2024.

\bibitem{pham2022split}
N.~D. Pham, T.~K. Phan, A.~Abuadbba, Y.~Gao, D.~Nguyen, and N.~Chilamkurti.
\newblock Split learning without local weight sharing to enhance client-side data privacy.
\newblock {\em arXiv:2212.00250}, 2022.

\bibitem{pham2024splitlearninglocalweight}
N.~D. Pham, T.~K. Phan, A.~Abuadbba, Y.~Gao, V.-D. Nguyen, and N.~Chilamkurti.
\newblock { Split Learning without Local Weight Sharing To Enhance Client-side Data Privacy }.
\newblock {\em IEEE TDSC}, (01):1--13, Apr. 5555.

\bibitem{podschwadt2021sok}
R.~Podschwadt, D.~Takabi, and P.~Hu.
\newblock Sok: Privacy-preserving deep learning with homomorphic encryption.
\newblock {\em arXiv:2112.12855}, 2021.

\bibitem{poirot2019split}
M.~G. Poirot, P.~Vepakomma, K.~Chang, J.~Kalpathy-Cramer, R.~Gupta, and R.~Raskar.
\newblock Split learning for collaborative deep learning in healthcare.
\newblock {\em arXiv:1912.04966}, 2019.

\bibitem{safeSplit2025rieger}
P.~Rieger, A.~Pegoraro, K.~Kumari, T.~Abera, J.~Knauer, and A.-R. Sadeghi.
\newblock Safesplit: A novel defense against client-side backdoor attacks in split learning.
\newblock In {\em NDSS}, 2025.

\bibitem{roth2022splitunet}
H.~R. Roth, A.~Hatamizadeh, Z.~Xu, C.~Zhao, W.~Li, A.~Myronenko, and D.~Xu.
\newblock Split-u-net: Preventing data leakage in split learning for collaborative multi-modal brain tumor segmentation, 2022.

\bibitem{rudin1992nonlinear}
L.~I. Rudin, S.~Osher, and E.~Fatemi.
\newblock Nonlinear total variation based noise removal algorithms.
\newblock {\em Physica D}, 1992.

\bibitem{ryan2011cloud}
M.~D. Ryan.
\newblock Cloud computing privacy concerns on our doorstep.
\newblock {\em Communications of the ACM}, 2011.

\bibitem{DISCO}
A.~Singh, A.~Chopra, V.~Sharma, E.~Garza, E.~Zhang, P.~Vepakomma, and R.~Raskar.
\newblock Disco: Dynamic and invariant sensitive channel obfuscation for deep neural networks, 2021.

\bibitem{tajalli2023feasibility}
B.~Tajalli, O.~Ersoy, and S.~Picek.
\newblock On feasibility of server-side backdoor attacks on split learning.
\newblock In {\em IEEE SPW}, 2023.

\bibitem{talaei2024adaptivedifferentialprivacyfederated}
M.~Talaei and I.~Izadi.
\newblock Adaptive differential privacy in federated learning: A priority-based approach.
\newblock {\em arXiv:2401.02453}, 2024.

\bibitem{thapa2022splitfed}
C.~Thapa, M.~A.~P. Chamikara, S.~Camtepe, and L.~Sun.
\newblock Splitfed: When federated learning meets split learning.
\newblock In {\em AAAI}, 2022.

\bibitem{practical2021}
T.~Titcombe, A.~J. Hall, P.~Papadopoulos, and D.~Romanini.
\newblock Practical defences against model inversion attacks for split neural networks.
\newblock In {\em ICLR Workshop on DPML}, 2021.

\bibitem{turina2021fsl}
V.~Turina, Z.~Zhang, F.~Esposito, and I.~Matta.
\newblock Federated or split? a performance and privacy analysis of hybrid split and federated learning architectures.
\newblock In {\em CLOUD}, 2021.

\bibitem{PrivateMail}
P.~Vepakomma, J.~Balla, and R.~Raskar.
\newblock Privatemail: Supervised manifold learning of deep features with differential privacy for image retrieval, 2021.

\bibitem{vepakomma2019reducing}
P.~Vepakomma, O.~Gupta, A.~Dubey, and R.~Raskar.
\newblock Reducing leakage in distributed deep learning for sensitive health data.
\newblock {\em arXiv:1812.00564}, 2019.

\bibitem{vepakomma2018split}
P.~Vepakomma, O.~Gupta, T.~Swedish, and R.~Raskar.
\newblock Split learning for health: Distributed deep learning without sharing raw patient data, 2018.

\bibitem{NoPeek}
P.~Vepakomma, A.~Singh, O.~Gupta, and R.~Raskar.
\newblock Nopeek: Information leakage reduction to share activations in distributed deep learning, 2020.

\bibitem{PSLF2023Wan}
X.~Wan, J.~Sun, S.~Wang, L.~Chen, Z.~Zheng, F.~Wu, and G.~Chen.
\newblock Pslf: Defending against label leakage in split learning.
\newblock In {\em ACM CIKM}, 2023.

\bibitem{wang2024stitch}
Y.~Wang, C.~Zhang, Z.~Zheng, J.~Wang, and X.~Li.
\newblock Stitch-able split learning assisted multi-uav systems.
\newblock {\em IEEE Open J. Comput. Soc}, 2024.

\bibitem{wu2024evaluating}
X.~Wu, H.~Yuan, X.~Li, J.~Ni, and R.~Lu.
\newblock Evaluating security and robustness for split federated learning against poisoning attacks.
\newblock {\em IEEE TIFS}, 2024.

\bibitem{xie2023label}
S.~Xie, X.~Yang, Y.~Yao, T.~Liu, T.~Wang, and J.~Sun.
\newblock Label inference attack against split learning under regression setting.
\newblock {\em arXiv:2301.07284}, 2023.

\bibitem{xu2024stealthy}
X.~Xu, M.~Yang, W.~Yi, Z.~Li, J.~Wang, H.~Hu, Y.~Zhuang, and Y.~Liu.
\newblock A stealthy wrongdoer: Feature-oriented reconstruction attack against split learning.
\newblock In {\em CVPR}, 2024.

\bibitem{yang2022differentially}
X.~Yang, J.~Sun, Y.~Yao, J.~Xie, and C.~Wang.
\newblock Differentially private label protection in split learning.
\newblock {\em arXiv:2203.02073}, 2022.

\bibitem{yu2023backdoor}
F.~Yu, L.~Wang, B.~Zeng, K.~Zhao, Z.~Pang, and T.~Wu.
\newblock How to backdoor split learning.
\newblock {\em Neural Networks}, 2023.

\bibitem{yu2024chronic}
F.~Yu, B.~Zeng, K.~Zhao, Z.~Pang, and L.~Wang.
\newblock Chronic poisoning: Backdoor attack against split learning.
\newblock In {\em AAAI}, 2024.

\bibitem{zeng2025gan}
B.~Zeng, S.~Luo, F.~Yu, G.~Yang, K.~Zhao, and L.~Wang.
\newblock Gan-based data reconstruction attacks in split learning.
\newblock {\em Neural Networks}, 2025.

\bibitem{zhang2024functionality}
L.~Zhang, X.~Gao, Y.~Li, and Y.~Liu.
\newblock Functionality and data stealing by pseudo-client attack and target defenses in split learning.
\newblock {\em IEEE TDSC}, 2024.

\bibitem{zhao2024splitaum}
K.~Zhao, X.~Chuo, F.~Yu, B.~Zeng, Z.~Pang, and L.~Wang.
\newblock Splitaum: Auxiliary model-based label inference attack against split learning.
\newblock {\em IEEE TNSM}, 2024.

\bibitem{zhao2018federated}
Y.~Zhao, M.~Li, L.~Lai, N.~Suda, D.~Civin, and V.~Chandra.
\newblock Federated learning with non-iid data.
\newblock {\em arXiv:1806.00582}, 2018.

\bibitem{zhu2019dlg}
L.~Zhu, Z.~Liu, and S.~Han.
\newblock Deep leakage from gradients.
\newblock In {\em NeurIPS}, 2019.

\bibitem{zhu2023passive}
X.~Zhu, X.~Luo, Y.~Wu, Y.~Jiang, X.~Xiao, and B.~C. Ooi.
\newblock Passive inference attacks on split learning via adversarial regularization.
\newblock {\em arXiv:2310.10483}, 2023.

\end{thebibliography}
